\newif\iffigs\figstrue

\documentstyle[12pt]{article}
\iffigs
  \input epsf
\else
\fi

\title{
{\hspace{265pt} \parbox{130pt}{\normalsize hep-th/9504061\\
               \normalsize ITPUWr 889\\
               \normalsize AMSPPS 199504-81-001}}\\
\ \\
Classical and quantum \\ massive string}
\author{Zbigniew Hasiewicz\\
{\small \it Institute of Theoretical Physics, Wroc{\l}aw University,
Wroc{\l}aw, Poland}\thanks{Institute of
Theoretical Physics, Wroc{\l}aw University, pl. Maxa Borna 9,
50-204 Wroc{\l}aw, Poland; E-mail: zhas@ift.uni.wroc.pl}
\and
Zbigniew Jask\'{o}lski\thanks{  Permanent address: Institute of
Theoretical Physics, Wroc{\l}aw University, pl. Maxa Borna 9,
50-204 Wroc{\l}aw, Poland; E-mail: jaskolsk@ift.uni.wroc.pl}\\
{\small \it
Institut de Recherche sur la Math\'{e}matique Avanc\'{e}e }\\
{\small \it
Universit\'{e} Louis Pasteur et CRNS, Strasbourg, France} }
\date{ April, 1995}
\begin{document}
\maketitle

\begin{abstract}

The classical and the quantum massive string model based on a modified
BDHP action is analyzed in the range of dimensions $1<d<25$. The
discussion concerning classical theory includes a  formulation of
the geometrical variational principle, a phase-space description of
the two-dimensional dynamics, and a detailed analysis of the target
space geometry of classical solutions. The model is quantized using
"old" covariant method. In particular an appropriate construction of
DDF operators is given and the no-ghost theorem is proved. For a critical
value of one of free parameters of the model the quantum theory acquires
an extra symmetry not present on the classical level. In this case the
quantum model is equivalent to the noncritical Polyakov string and to
the old Fairlie-Chodos-Thorn massive string.

\end{abstract}

\section{Introduction}

The first quantum model of massive string was
proposed  twenty years ago by
Chodos and Thorn
\cite{ct}. It was originally formulated as the tensor product of
$d$-copies of
the standard free field representation of the Virasoro algebra and
one copy of the free field Fairlie representation with the central charge
$c =26-d$. Although satisfying all consistency conditions of
a free string model it had some important drawbacks.
First of all it contained a non physical
continuous internal quantum number.
Secondly because of the lack of the path
integral representation  it was not clear how to implement
the idea of "joining-splitting" interaction. The attempts to
construct string amplitudes by means of the operator formalism known from
 the critical dual models lead to tree amplitudes with
a continuous range of intercepts.  Finally rather ad hoc
phase space formulation  made the
 physical  interpretation of the classical system obscure.
For these reasons the FCT string did not received much attention
at that time.

Soon after the Polyakov paper
\cite{pol} on conformal anomaly in the
BDH string \cite{bdh},
the Fairlie realization of the Virasoro algebra reappeared  in the
 quantization of Liouville theory \cite{ctgn}.
One of the early attempts
to clarify
the relation between noncritical string and the Liouville theory
suggested by Polyakov's work,
was made by Marnelius \cite{marnel}. He
considered  a modified string model determined by the world
sheet action
\begin{eqnarray}
S[M,g,\varphi,x] &=&
-{\alpha \over 2\pi} \int\limits_M
\sqrt{-g}\,d^2z\, g^{ab}\partial_a x^{\mu}
\partial_bx^{\nu}\eta_{\mu\nu}
\label{ctp0} \\
&-& {\beta \over 2\pi} \int\limits_M \sqrt{-g}\,d^2z\, \left(  g^{ab}
\partial_a \varphi \partial_b \varphi + 2 R_g\varphi \right)
- \mu \int\limits_{ M} \sqrt{-g}\,d^2z\, \exp (\varphi)
\;\;\;.\nonumber
\end{eqnarray}
The  analysis of \cite{marnel} was devoted to the case of
non-vanishing cosmological constant. As a side remark
it was pointed out
that for $\mu = 0$ and $\beta = {25-d\over 48}$
the canonically quantized model
 should be equivalent  with the FCT string \cite{ct}.

The action (\ref{ctp0}) regarded as
a two-dimensional conformal field theory
action  describes
 a special case of the induced (Liouville) 2-dim gravity
coupled to the conformal matter. This  system
has been extensively studied over last few years, both as a
noncritical dual model \cite{ddk}
(this application being restricted by the famous
$c=1$
barrier) and as a dilaton gravity toy model
for analyzing formation and evaporation of black holes \cite{black}.
(For a review with references to the
literature see \cite{gimo}.)

The basic idea of the present paper is to consider the action
(\ref{ctp0}) with $\mu= 0$
as a world sheet action of  relativistic one-dimensional
extended object,
rather than a  conformal field theory input of the general
dual model construction.   The main result is
that such action leads in the range of dimensions $1<d<25$
to a consistent  classical and quantum relativistic string theory.
Since for all admissible values of $\beta$ parameter
physical states on first excited level are massive we call
this model the massive string.
Our motivation to consider it stems from
the recent analysis  of noncritical Polyakov string \cite{jame}, where
it was shown that the Polyakov sum over random surfaces leads
in the range $1<d<25$ to
a consistent quantum mechanics of one-dimensional relativistic
objects, equivalent to the old FCT string model.
One
can expect that  the same quantum theory can be obtained
 from the modified BDHP action (\ref{ctp0}),  by standard quantization
techniques.

One of the results of the present paper is that for a critical value of
the parameter $\beta$ these two procedures are indeed equivalent.
This  yields a new insight into
the symmetry structure of the noncritical Polyakov  string and
gives some new hints for constructing a consistent joining-splitting
interaction in this model. Although the problem of interacting
noncritical string was our main motivation for analyzing
the free massive string, the model is very interesting by its own.
Both the classical and the quantum theory exhibit  nontrivial
structures and rise many interesting physical and geometrical
questions.

The organization of the paper is as follows. In Section 2
 we formulate a geometrical
variational  principle for the massive string. In order to emphasize the
geometrical character of the theory we use reparameterization
invariant boundary
conditions. This clarifies the geometrical origin of the constraint
removing the continuous
internal degree of freedom from the model.
It also considerably simplifies
 the phase-space analysis
of the two-dimensional dynamics of the system which is given in
Section 3.

The target space interpretation of the classical
model as well as its relation to  the notion of classical
causality  is analyzed in Section 4.
It is shown in particular that if
one assumes
the strong classical causality condition then the massive string
coincides with the Nambu-Goto and the BDHP classical models.
It turns out however that the  model
admits "stringy" interpretation with
a weaker notion of causality, which allows for an
essentially wider space
of classical solutions.
In this respect the massive string can be seen as an extension of
the classical string  models.

One of important features of
the classical massive string is that it does not admit the target
space light-cone gauge.  Due to the nonzero central term in the
classical (Poisson bracket) algebra
$$
\{{L}_m,{L}_n \} = i(m-n){L}_{n+m}
-4\beta im^3\delta_{m,-n}\;\;\;.
$$
the  constraints  are of the second kind.
Thus, in contrast to the BDHP and the Nambu-Goto string models,
  solving  constraints before quantization is a
prohibitively difficult task.

In Section 5 we use the covariant operator method to quantize
the model. Following  Brower's ideas we construct the DDF operators
and show that they generate the full
space of physical states. The metric structure on this
space is then completely determined by general results concerning
unitary irreducible highest weight representations \cite{gaw} of the
Virasoro algebra. The corresponding no-ghost theorem is
formulated for the whole range of free parameters of the model.

Section 6 contains conclusions and a brief discussion of some
open problems.

\section{Variation  principle}

A classical open CTP string trajectory will be described by
a set $(M,g,\varphi,x)$ where
$M$ is a rectangle-like 2-dim oriented manifold with distinguished
"initial" $\partial_iM$ and "final" $\partial_fM$ opposite boundary
components,
$g$ is a pseudo-Riemannian metric on $M$,
$\varphi : M \rightarrow {\bf R}$ is a scalar function on $M$ and
$x: M \rightarrow {\bf R}^d$ is a map from $M$ into d-dim Minkowski
space.
It is assumed that the metric $g$ on $M$  has a trivial
causal structure such that the "initial" and "final" boundary
components are space-like while the other components $\partial
M\setminus
(\partial_iM \cup \partial_fM)$ are time-like.

On the space of classical trajectories we consider the action
functional
\begin{equation}
\parbox{300pt}{
\begin{eqnarray*}
S[M,g,\varphi,x] &=&
-{\alpha \over 2\pi} \int\limits_M
\sqrt{-g}\,d^2z\, g^{ab}\partial_a x^{\mu}
\partial_bx^{\nu}\eta_{\mu\nu}
 \\
&-& {\beta \over 2\pi} \int\limits_M \sqrt{-g}\,d^2z\, \left(  g^{ab}
\partial_a \varphi \partial_b \varphi + 2 R_g\varphi \right)
\;\;\;,
\end{eqnarray*} }
\label{v1}
\end{equation}

where $R_g$ is the scalar curvature of $g$,
 and $\eta^{\mu\nu} = {\rm diag} (-1,+1,...,+1)$.
The constant $\beta$ is dimension-less in units of $\hbar$, and the
constant
$\alpha$ is conventionally expressed in terms of the slope parameter
$\alpha'$ with dimension of length-squared
$
\alpha = {1 \over 2\alpha'}
$.

The group of global symmetries of the action functional (\ref{v1})
consists
of Poincare transformations in the target space and constant
rescalings
of the internal metric $g$.

The action  is invariant with respect to
general diffeomorphisms
$f:M\rightarrow M'$ preserving the initial and final boundary
components and
their orientations.
The transformation rule for the scalar curvature
$$
R_{{\rm e}^{\varrho}g} = {\rm e}^{-\varrho} ( R_{g}  + \Box_g  \varrho )
\;\;\;,
$$
implies that it is also invariant with
respect to the rescalings of metrics $g \rightarrow {\rm e}^{\varrho}g$
with a conformal factor
$\varrho$ satisfying the equation
$$
\Box_g \varrho \equiv -{1\over \sqrt{-g}}\partial_a(\sqrt{-
g}g^{ab}\partial_b \varrho)\;=\;0  \;\;\;.
$$

Before formulating variational  principle it
 is convenient to restrict this huge gauge invariance
by introducing some partial gauge fixings.
This will be done in two steps, first
partially restricting reparameterization
invariance
and then introducing gauge fixing for the rescaling
symmetry.

Let us fix a model manifold $M$ and a normal direction $n$ along the
boundary
$\partial M$.  For any string trajectory $(N,g,\varphi,x)$
there exists a diffeomorphism $f: M \rightarrow N$
such that the normal direction $n_{f^*g}$ of the metric $f^*g$
is proportional to $n$, and the boundary components meet orthogonally
 at corners  with respect to $f^*g$.
 This means that fixing $M$ and imposing
these conditions on metrics provides a good partial gauge fixing for
the reparameterization invariance, which we shall call the $(M,n)$-gauge.
 In this gauge  the space of classical trajectories
is the Cartesian product ${\cal M}^n_M \times {\cal W}_M \times {\cal
E}^d_M$
where ${\cal M}^n_M$ is the space of metrics with the normal
direction $n_g \propto n$ and with right angles at the corners,
${\cal W}_M$ is the space of real-valued
scalar functions on $M$, and ${\cal E}^d_M$ is the space of maps from
$M$ into
d-dim Minkowski target space. The reparameterizations form the group
${\cal D}^n_M$ of diffeomorphisms
of $M$ preserving corners and the normal direction $n$. The action of
${\cal D}^n_M$ on
 ${\cal M}^n_M \times {\cal W}_M \times {\cal E}^d_M$ is given by
\begin{equation}
\label{v2}
(g,\varphi, x)
{ \begin{picture}(60,20)(0,0) \put(5,3){\vector(1,0){50}}
\put(15,10){$\scriptstyle
f \in   {\cal D}_M^n
$ }
\end{picture}}
(f^*g,\varphi\circ f, x\circ f)
\;\;\;.
\end{equation}

In the $(M,n)$-gauge the rescalings also form
a group which can be identified with the abelian group
 ${\cal K}_{\partial M}$
of 1-forms on $\partial M$ satisfying
$
\int \widetilde{\kappa} = 0
$.
For any  $g \in {\cal M}_M^n$ and $\widetilde{\kappa} \in
{\cal K}_{\partial M}$ there exists a unique solution
 $\varrho[g,\widetilde{\kappa}]$ to
the boundary problem
$$
\Box_g  \varrho \;=\;0\;\;\;\;,
\;\;\;\;
(n^a_g \partial_a \varrho) e^g \;=_{|\partial M}\; \widetilde{\kappa} \;\;\;,
$$
satisfying
the normalization condition
$$
\int\limits_M \sqrt{-g}\, d^2 z\, \varrho= 0
\;\;\;.
$$
$e^g$ in the boundary condition above denotes the einbein
(or the volume form) of the
1-dim metric induced on $\partial M$.
The action of
  ${\cal K}_{\partial M}$ on the space of trajectories
${\cal M}_M^n \times {\cal W}_M \times {\cal E}^d_M$ is given by
\begin{equation}
\label{v3}
 (g,\varphi, x)
{ \begin{picture}(60,20)(0,0) \put(5,3){\vector(1,0){50}}
\put(15,10){$\scriptstyle
\widetilde{\kappa} \in   {\cal K}_{\partial M}
$ }
\end{picture}}
({\rm e}^{\varrho[g,\widetilde{\kappa}]}g,\varphi, x)
\;\;\;.
\end{equation}

It follows that in the $(M,n)$-gauge the group of gauge transformations
is a semi-simple product of  ${\cal D}_M^n$ and ${\cal K}_{\partial M}$.
Let us observe that the transformation rule for the geodesic
curvature $\kappa_g$ of the boundary $\partial M$
$$
\kappa_{{\rm e}^{\varrho}g} = {\rm e}^{-{\varrho\over 2}}
(\kappa_g + {\textstyle {1\over 2}} n^a_g \partial \varrho) \;\;\;,
$$
implies that for a fixed
$\widetilde{\kappa} \in   {\cal K}_{\partial M}$
the condition
$
\kappa_g e^g= \widetilde{\kappa}
$
is a good gauge fixing for the gauge symmetry (\ref{v3}).
Choosing  $\widetilde{\kappa}=0$ one gets the
${\cal D}^n_M$-invariant gauge fixing condition
\begin{equation}
\label{v4}
\kappa_g =0\;\;\;,
\end{equation}
which we shall  call the geodesic gauge. It can be thought of as a
counterpart of the constant curvature gauge in the BDHP model.
In this partial gauge
the  space of trajectories is given by
the product
${\cal M}^{n0}_M \times {\cal W}_M \times {\cal E}^d_M$
where  ${\cal M}^{n0}_M $ is the space of metrics from
${\cal M}^{n}_M$ with zero geodesic curvature.
Let us note that the same space of metric has been obtained in the
Polyakov sum over bordered surfaces  from
the requirement of a proper structure of boundary conditions for the
Faddeev-Popov operator \cite{ja}.
In the geodesic gauge the  group of
 gauge transformations  reduces to
${\cal D}^n_M$ and its action on the space of trajectories is simply
given by the restriction of (\ref{v2}) to
${\cal M}^{n0}_M \times {\cal W}_M \times {\cal E}^d_M$.

Calculating the full variation of (\ref{v1}) on
${\cal M}^{n0}_M \times {\cal W}_M \times {\cal E}^d_M$ one gets
\begin{eqnarray}
-\delta S[M,g,\varphi,x] &=& {\alpha\over \pi} \int\limits_M
\sqrt{-g}\,d^2z\, (\Box_g x^{\mu} ) \delta x_{\mu}
+ {\alpha\over \pi}
\int\limits_{\partial M} ds\, n^a\partial_a x^{\mu}\delta x_{\nu}
      \nonumber\\
&+& {\beta\over \pi} \int\limits_m \sqrt{-g}\,d^2z\, \left(
\Box_g \varphi + R_g \right)\delta \varphi
+ {\beta\over \pi} \int\limits_{\partial M} ds\,
n^a\partial_a \varphi  \delta \varphi
\label{v5} \\
&+&  {1\over 2} \int\limits_M \sqrt{-g} \,d^2z\, \left[
{\alpha\over \pi} \left( \partial_ax^{\mu} \partial_bx_{\mu}
-{1\over 2}g_{ab}g^{cd}\partial_cx^{\mu} \partial_dx_{\mu} \right)
\right.\nonumber\\
& &\;\;\;+\; {\beta\over \pi}
\left.\left(\partial_a\varphi \partial_b\varphi
-{1\over 2}g_{ab}g^{cd}\partial_c\varphi \partial_d\varphi
-2\Box_g\varphi g_{ab} -2\nabla_a\nabla_b\varphi \right)
\right]\delta g^{ab}
\nonumber\\
&+& {\beta\over\pi}
 \int\limits_{\partial M_s} ds\, t^at^b\delta
g_{ab}n^c\partial_c\varphi
-\;{\beta\over\pi} \int\limits_{\partial M_t} ds\, t^at^b\delta
g_{ab}n^c\partial_c\varphi
\;\;\;,\nonumber
\end{eqnarray}
where
$\partial M_t$, $\partial M_s$ denote the time-like and the space-
like
boundary components, respectively.
Let us note that in the geodesic gauge
 variations in the metric sector satisfy the boundary conditions
\begin{eqnarray}
n^at^b\delta g_{ab}& = &0\;\;\;,\label{v6}\\
n^a\nabla_a ( g^{bc}\delta_{bc} ) - n^a\nabla^b \delta g_{ab}
&=&0\;\;\;,\nonumber
\end{eqnarray}
along all boundary components.

The requirement of the vanishing bulk variation leads to the following
equations of motion:
\begin{eqnarray}
\label{v7}
\Box_g x^{\mu}& =& 0 \;\;\;,\\
\label{v8}
\Box_g \varphi + R &=& 0 \;\;\;,\\
\label{v9}
T_{ab} &=&0\;\;\;,
\end{eqnarray}
where $T_{ab}$ is the energy-momentum tensor:
\begin{eqnarray}
\nonumber
{T_{ab}}   &=& {\alpha\over\pi} \left(
\partial_ax^{\mu} \partial_bx_{\mu}
- {1\over 2} g_{ab} g^{cd} \partial_c x^{\mu} \partial_d
x_{\mu}
\right) \\
&+ & {\beta\over\pi} \left( \partial_a\varphi \partial_b\varphi
-{1\over 2}g_{ab}g^{cd}\partial_c\varphi \partial_d\varphi
-2\Box_g\varphi g_{ab} -2\nabla_a\nabla_b\varphi \right)\;\;\;.
\nonumber
\end{eqnarray}
The trace part of (\ref{v9}) yields
\begin{equation}
\label{v10}
\Box_g \varphi = 0\;\;\;,
\end{equation}
which, together with the equation (\ref{v8}) implies
\begin{equation}
\label{v11}
R_g = 0\;\;\;.
\end{equation}

In order to obtain a well posed variational  problem one has to impose
boundary conditions in the $x$- and $\varphi$-sector
for which the boundary terms in
the expression (\ref{v5}) vanish.  Since the classical system
under consideration is supposed to describe a free open string with
some additional internal structure the variations $\delta x^{\mu}$ and
$\delta\varphi$
 should be arbitrary at the "ends". This
leads to the following boundary conditions  along $\partial M_t$:
 \begin{equation}
 \label{v12}
n^a\partial_a x^{\mu}\; =\; 0\;\;\;,
\;\;\;n^a\partial_a \varphi \;=\;0\;\;\;.
\end{equation}

On the space-like boundary components one needs Dirichlet type
boundary
conditions describing initial and final configurations of the system.
  ${\cal D}_M^n$-invariant boundary conditions of this
type
were first introduced in \cite{dop}.
More recently these boundary conditions have been
analyzed in the context of  the covariant path integral
quantization of
noncritical Polyakov string \cite{jame}. The derivation
of the gauge invariant boundary condition in the $x$- and $\varphi$-
sector
of the present model
is essentially the same. The basic idea is to impose the Dirichlet
boundary
conditions in the parameterization
in which the  einbeins induced on the
space-like boundary components are constant.

For every classical trajectory $(g,\varphi,x) \in
 {\cal M}^n_M \times {\cal W}_M \times {\cal E}^d_M$ we define
 the initial configuration as a set $(e^g_i,\varphi_i,x_i)$
 where
 \begin{equation}
\label{v13}
 (e^g_{i})^2\; =\; g_{ab}t^at^b(dt)^2_{|\partial M_i}\;\;\;,\;\;\;
 \varphi_i \;=\; \varphi_{|\partial M_i}\;\;\;,
 \;\;\;
 x^{\mu}_i \;= \; x^{\mu}_{|\partial M_i}\;\;\;.
 \end{equation}
All possible initial configurations form the space
$$
{\cal P}_i \equiv {\cal M}_i \times {\cal W}_i \times {\cal E}^d_i
\;\;\;,
$$
where ${\cal M}_i$ consists of all einbeins on $\partial M_i$,
${\cal W}_i$ is the space of real-valued functions on
$\partial M_i$,
and ${\cal E}^d_i$ is the space of maps
$x_i : \partial M_i \rightarrow {\bf R}^d $. Due to
the boundary conditions (\ref{v12}) the
functions from  ${\cal W}_i$ and $ {\cal E}^n_i$
satisfy Neumann boundary conditions at the ends of
$\partial M_i$.

The initial configuration defined by (\ref{v13}) is
${\cal D}_M^n$-covariant, i.e. for $f \in {\cal D}_M^n$ one has
$$
(e^{f^*g}_i,(\varphi\circ f)_i,(x\circ f)_i) = ( \gamma_i^*e^g_i,
\varphi_i\circ\gamma_i,x_i\circ\gamma_i)\;\;\;,\;\;\;
\gamma_i = f_{|\partial M_i}\;\;.
$$
Due to this property one can
 define the space ${\cal C}_i$ of gauge invariant initial
configurations as the quotient
$$
{\cal C}_i \equiv \frac{{\cal M}_i \times {\cal W}_i \times {\cal
E}^d_i}{
{\bf R}_+\times {\cal D}_i}\;\;\;,
$$
where ${\cal D}_i$ is the group of orientation preserving
diffeomorphisms
of the initial boundary component $\partial M_i$, and the action of
${\bf R}_+\times {\cal D}_i$ on
${\cal M}_i \times {\cal W}_i \times {\cal E}^d_i$
is given by
\begin{equation}
\label{v14}
 (e_i,\varphi_i,x_i)
{ \begin{picture}(82,20)(0,0) \put(5,3){\vector(1,0){72}}
\put(13,10){$\scriptstyle
(\widetilde{\lambda},\gamma) \in {\bf R}_+ \times  {\cal D}_i
$ }
\end{picture}}
(\widetilde{\lambda}\gamma^*e_i,\varphi_i \circ \gamma, x_i \circ
\gamma) \;\;.
\end{equation}

The group ${\bf R}_+$ of constant rescalings of
einbeins has been introduced  in order to avoid over complete boundary
conditions.
In fact using the Gauss-Bonnet
theorem one can easily show that for the zero-curvature metrics from
${\cal M}^n_M$ with fixed geodesic curvature of $\partial M$
the internal length of the boundary
cannot be arbitrary.

The space of gauge invariant final configurations
${\cal C}_f$
is defined in a similar way. It is convenient to use a common
description for both spaces introducing a model interval $L$ and
the quotient space
$$
{\cal C}_L \equiv \frac{{\cal M}_L \times {\cal W}_L \times {\cal
E}^d_L}{
{\bf R}_+\times {\cal D}_L}\;\;\;
$$
canonically isomorphic with ${\cal C}_i$ and ${\cal C}_f$.
If we assume the induced orientations on $\partial M_i$ and
$\partial M_f$, the isomorphisms are given  by
\begin{eqnarray*}
{\cal C}_L \ni [(\widetilde{e},
\widetilde{\varphi},\widetilde{x})]& \longrightarrow  &
[(\gamma^*_i\widetilde{e},\widetilde{\varphi}\circ \gamma_i ,
\widetilde{x}\circ\gamma_i)] \in {\cal C}_i
\;\;\;,\\
{\cal C}_L \ni [(\widetilde{e},
\widetilde{\varphi},\widetilde{x})]& \longrightarrow  &
[(\gamma^*_f\widetilde{e},\widetilde{\varphi}\circ \gamma_f ,
\widetilde{x}\circ\gamma_f)] \in {\cal C}_f
\;\;\;,
\end{eqnarray*}
where $\gamma_i : \partial M_i \rightarrow L$ is an arbitrary
orientation preserving diffeomorphism,
$\gamma_f : \partial M_f \rightarrow L$ is an arbitrary
orientation reversing diffeomorphism,  and the square brackets denote
the
gauge orbits of corresponding elements of ${\cal P}_i$, ${\cal
P}_f$, and ${\cal P}_L$.

The gauge invariant boundary conditions for classical trajectory
$(g,\varphi,x)\in {\cal M}^n_M\times {\cal W}_M\times {\cal E}^d_M$
are then defined by
\begin{equation}
\label{v15}
[(e_i,\varphi_i,x_i,\kappa_i)]\;=\;c_i\;\;\;,
\;\;\;[(e_f,\varphi_f,x_f,\kappa_f)]\;=\;c_f
\;\;\;,
\end{equation}
where $c_i,c_f \in {\cal C}_L$.

 We shall check whether the variational  problem (\ref{v5})
is well posed with the boundary conditions (\ref{v15}). For this
purpose it is convenient to introduce a more tractable description of
the space of gauge invariant configurations.
Let ${\widehat{e}}$ be an einbein on the model interval $L$. One
can easily check that for every
$({\bf R}_+\times{\cal D}_L)$-orbit $c \in {\cal C}_L$ there exists
a unique $(\widetilde{\varphi},\widetilde{x})
\in {\cal W}_L \times {\cal E}^d_L $
such that $c = [(\widehat{e},\widetilde{\varphi},\widetilde{x})]$.
In fact the condition $\widetilde{e} = \widehat{e}$ defines a smooth
gauge slice for the action (\ref{v14}),
and provides 1-1 parametrization of  ${\cal C}_L$
$$
 {\cal W}_L \times {\cal E}^d_L
 \ni (\widetilde{\varphi},\widetilde{x})
 \longrightarrow [(\widehat{e},\widetilde{\varphi},\widetilde{x})]
\in {\cal C}_L\;\;\;.
$$
Let $c_i = [(\widehat{e},\widetilde{\varphi}_i,\widetilde{x}_i,
\widetilde{\kappa}_i)]$
and $c_f = [(\widehat{e},\widetilde{\varphi}_f,\widetilde{x}_f,
\widetilde{\kappa}_f)]$.
Then the boundary conditions (\ref{v15}) take the form
\begin{eqnarray}
x_i\;=\;\widetilde{x}_i \circ \widehat{\gamma}_i[g]\;\;\;&,&
\;\;\;\varphi_i\;=\;\widetilde{\varphi}_i \circ
\widehat{\gamma}_i[g]\;\;\;,
\label{v16}\\
x_f\;=\;\widetilde{x}_f \circ \widehat{\gamma}_f[g]\;\;\;
&,&
\;\;\;\varphi_f\;=\;\widetilde{\varphi}_f \circ
\widehat{\gamma}_f[g]\;\;\;,
\nonumber
\end{eqnarray}
where the diffeomorphisms
$$
\widehat{\gamma}_i[g] : \partial M_i \longrightarrow L\;\;\;,
\;\;\;\widehat{\gamma}_f[g] : \partial M_f \longrightarrow L\;\;\;
$$
are uniquely determined by the equations
\begin{equation}
\label{v17}
\widehat{\gamma}_i[g]^*\widehat{e}\;\propto\;e_i\;\;\;,
\;\;\;\widehat{\gamma}_f[g]^*\widehat{e}\;\propto\;e_f\;\;\;,
\end{equation}
and the convention concerning orientation.

The analysis of the boundary conditions (\ref{v16}) is the same for
both
space-like boundary components and we restrict our considerations to
the
initial boundary. Let $s\in [0,1]$ be an arbitrary parameterization of
$\partial M_i$. The variation of (\ref{v16}) yields
\begin{equation}
\label{v18}
\delta x_i(s)\;=\;(\partial_s x_i)(s)\delta
\widetilde{\gamma}_i[g](s)\;\;\;,\;\;\;
\delta \varphi_i(s)\;
=\;(\partial_s \varphi_i)(s)\delta \widetilde{\gamma}_i[g](s)\;\;\;
\end{equation}
where
$$
\delta \widetilde{\gamma}_i[g](s) =
\frac{\delta
\widehat{\gamma}_i[g](s)}{\partial_s\widehat{\gamma}_i[g](s)}\;\;\;.
$$
In the parameterization $\widehat{s} \in [0,1]$ of $L$ such that
$ \widehat{e}
= {\rm const}\cdot d\widehat{s}$ the solution to the equation
(\ref{v17}) can be easily calculated:
$$
\widehat{\gamma}_i[g](s) = {1\over l_i} \int\limits_0^s e_i (s')
ds'\;\;\;,
$$
where
$e_i(s) = \sqrt{g(\partial_s,\partial_s)}$ and
$l_i = \int\limits_0^1 e_i(s)ds$. Using the equation above one gets
$$
\delta \widetilde{\gamma}_i[g](s) = -\delta l
\frac{\widehat{\gamma}_i[g](s)}{e_i(s)} + {1\over e_i(s)}
\int\limits_0^s \delta e_i(s') ds'
$$
and
\begin{equation}
\label{v19}
\frac{\delta g(\partial_s,\partial_s)}{
\sqrt{g(\partial_s,\partial_s)}}
= 2\partial_s(e_i \delta \widetilde{\gamma}_i[g]) +
2{\delta l_i\over l_i} e_i\;\;\;.
\end{equation}
Inserting (\ref{v18}), and (\ref{v19}) into the
$\partial M_i$-boundary terms in (\ref{v5}) one obtains
\begin{eqnarray*}
{1\over\pi}\int\limits_{\partial M_i} e_i\!\!\!& ds&\!\!\!
\left[ \alpha n^a\partial_a x^{\mu}\delta x_{\mu} +
\beta n^a\partial _a \varphi\delta \varphi +
\beta t^at^b\delta g_{ab} n^c\partial_c \varphi \right] \;=\\
&=&
{1\over\pi}\int\limits_0^1 e_i\, ds\,
\left[ \alpha n^a\partial_a x^{\mu}\partial_s x_{\mu} +
\beta n^a\partial_a \varphi\partial_s \varphi -
2\beta \partial_s( n^a\partial_a \varphi) \right]
\delta \widetilde{\gamma}_i[g] \\
&+& {\delta l_i\over l_i}{2\beta\over\pi} \int\limits_0^1 e_i(s)\,ds\,
n^a\partial_a\varphi\;\;\;.
\end{eqnarray*}
Note that for the metric variations $\delta g$ satisfying the
conditions  (\ref{v6})
the variations $\delta \widetilde{\gamma}_i[g](s)$, $\delta l_i$
are arbitrary and independent. It follows that
the $\partial M_i$--boundary terms vanish if and only if
the following conditions are satisfied
\begin{equation}
\label{v20}
\alpha n^at^b\partial_a x^{\mu}\partial_b x_{\mu} +
\beta n^a t^b \partial_a \varphi \partial_b \varphi +
2\beta n^at^b \nabla_a\nabla_b \varphi\;=\;0
\;\;\;,
\end{equation}
\begin{equation}
\label{v21}
\int\limits_{\partial M_i} e_i \, ds \, n^a\partial_a \varphi
\;=\;0\;\;\;,
\end{equation}
where in the derivation of (\ref{v20}) the identity
$
n^at^b \nabla_a\nabla_b \varphi = \kappa_g t^a\partial_a \varphi -
t^a\partial_a(n^b\partial_b \varphi)
$
and the condition $\kappa_g =0$ have been used.

The conditions (\ref{v20},\ref{v21}) should be regarded as (off-
shell)
constraints on
possible boundary values of dynamical variables.
The first condition implies that a part of the Euler-Lagrange
equations (\ref{v9}) --
the $T_{nt}$
component
of the energy-momentum tensor along the space-like boundary --
should be regarded as an off-shell condition.  The second condition
 means
that the zero mode of the momenta of $\varphi$ should be zero
off-shell. This
 removes the unwanted continuous internal degree of freedom and
is very important for the "stringy" interpretation of the model.
Since the both conditions are consistent with the  equations of motions
the variational  problem is  well posed.

 \section{Two-dimensional dynamics}

  In string models one has two conceptually different notions
  of evolution:
 the inner time evolution, and the  evolution
 in the target space. Although in some cases
 (e.g. Nambu-Goto string in the light-cone gauge) the inner and the
 target times can be identified, the interpretation of
 classical trajectories of two-dimensional system
 as histories
 of a  one-dimensional extended object in the target space, is a
 difficult problem and
 has to be analyzed in each model separately.
 In this section we shall concentrate on the  two-dimensional
 dynamics of the massive string  model,
 leaving the discussion of the target space evolution
 to the next section.

 In general
 the space of states and the time evolution
 of a classical system are  given
 in terms of Cauchy data for the Euler-Lagrange equations of the
 corresponding
 variational  principle. This formulation of dynamics assumes the
 existence
 of an evolution parameter for which the Cauchy problem
 is well posed.
 In the case of systems with reparameterization invariance there is
 in general  no
 gauge independent notion of (inner) time.
 The standard method of dealing with this problem is to  formulate the
variational
 principle with some fixed choice of time parameter.
 In this formulation  the initial and final boundary conditions are
 gauge dependent.
 The structure of the Cauchy data for the resulting Euler-Lagrange
 equations can be  analyzed by the
 phase space Dirac method.
 If the Hamiltonian of the classical system obtained in this way
 weakly vanishes one gets a consistent formulation independent
 of the choice of inner time. The classical system determined
 by the action functional (\ref{ctp0}) has been analyzed within this
 framework in a number of papers \cite{phas}.

 In contrast to the  scheme  above the formulation
 of the variational  problem given in the previous section is
 gauge invariant.
 In consequence
 the problem is well defined on the quotient space
 \begin{equation}
 \label{ctp23}
 \frac{ {\cal M}^{n0}_M \times {\cal W}_M \times {\cal E}^d_M}{ {\cal
 D}^n_M}
 \;\;\;.
 \end{equation}
 For a given choice of boundary conditions the solution $(g,\varphi,x)$
  of the Euler-Lagrange equations is determined up to the
 ${\cal D}^n_M$-action and can be seen as a point in the space
 (\ref{ctp23}). The interpretation of the Euler-Lagrange equations
 as dynamical equations requires an introduction of an evolution
 parameter.
 The uniformization
of metrics from
 ${\cal M}^{n0}_M$ allows for a ${\cal D}^n_M$-invariant
 definition of  the  inner time as
 the corresponding Teichm\"{u}ller parameter.
  As we shall see this leads to a consistent phase space
 formulation. The advantage of the gauge invariant description
 of the 2-dim dynamics is that it minimalizes the number of
 dynamical variables and therefore the number of  constraints
 in the corresponding phase space formulation. Moreover the
 dynamical variables, the constraints, and the equations
 involved have a clear geometrical
 interpretation.

 In order to analyze the dynamical content of the model one
needs some parametrization of the quotient space (\ref{ctp23}).
With our  choice of the inner time
the  conformal gauge is especially
 convenient:
 \begin{eqnarray}
 \label{ctp24}
 g& =& {\rm e}^{\varrho} \widehat{g}_t\;\;\;,\\
(\widehat{M}_t,\widehat{g}_t)& = &\left([0,t]\times[0,\pi],
  \left( \parbox{19pt}{ \scriptsize
 -1 \makebox[1pt]{}  0 \\
 \makebox[2pt]{}0 \makebox[1pt]{} 1 }
  \right)\right) \;\;\;,\;\;\; t\in{\bf R}_+\;\;\;.
                              \nonumber
  \end{eqnarray}
 In this gauge each point in the quotient (\ref{ctp23}) is uniquely
 represented by
 a set $(t,\varrho,\varphi,x) \in {\bf R}_+\times {\cal W}_M \times
 {\cal W}_M
 \times {\cal E}^d_M$.

In the conformal gauge the action functional (\ref{v1}) takes
 the form
 \begin{equation}
 \label{ctp25}
 S[t,\varrho,\varphi,x] =
  \int\limits_0^t d\tau \int\limits_0^{\pi} d\sigma
 \left[{\alpha \over 2}\left(  \dot{x}^2 -  x'^2 \right)+
 {\beta \over 2}
 \left(  \dot{\varphi}^2 - \varphi'^2  +2\dot{\varrho}\dot{\varphi}
 -2\varrho'\varphi' \right)\right]\;\;\;,
 \end{equation}
 where dot and prime stand for the partial derivatives with respect
 to the parameters $\tau$ and $\sigma$ respectively.
 By simple calculations one gets
 the equations of motions
 (\ref{v7},\ref{v10},\ref{v11})
 \begin{eqnarray}
 -\ddot{x}^{\mu} + {x^{\mu}}'' &=&0
 \;\;\;,\nonumber\\
 -\ddot{\varphi} + \varphi'' &=&0
 \;\;\;,\nonumber\\
 -\ddot{\varrho} + \varrho'' &=&0
 \;\;\;,      \label{td3}
 \end{eqnarray}
 and the energy momentum tensor
 \begin{eqnarray}
  T_{\tau \tau} &=&
 {\alpha\over 2\pi}
\left(  \dot{x}^2 + x'^2 \right)
+ {\beta \over 2\pi}
\left( \dot{\varphi}^2 + \varphi'^2 \right)
+ {\beta\over\pi}\dot{\varrho}\dot{\varphi}
+ {\beta\over\pi}\varrho'\varphi'
-2{\beta\over\pi} \ddot{\varphi}
\;\;\;,\nonumber\\
 T_{\sigma \sigma} &=&
 {\alpha\over 2\pi}
\left(  \dot{x}^2 + x'^2 \right)
+ {\beta \over 2\pi}
\left( \dot{\varphi}^2 + \varphi'^2 \right)
 + {\beta\over\pi}\dot{\varrho}\dot{\varphi}
+ {\beta\over\pi}\varrho'\varphi'
-2{\beta\over\pi} \varphi''
\;\;\;,\label{td4a}\\
 T_{\sigma \tau}&=& T_{\tau \sigma } \;=\;
 {\alpha\over\pi}\dot{x}x' + {\beta\over\pi}(\dot{\varphi}\varphi' +
\dot{\varrho}\varphi' + \varrho'\dot{\varphi} - 2\dot{\varphi}')\;\;\;.
 \label{td4b}
 \end{eqnarray}
 The boundary conditions along the time-like boundary components are given
 by
 \begin{equation}
 \label{td5}
 {x^{\mu}}' \;=\;0\;\;\;,\;\;\;\varphi'\;=\;0\;\;\;,\;\;\;\varrho'\;=\;0
 \;\;\;.
 \end{equation}
 Using natural identification of $\partial M_i$ and $\partial M_f$
with the model interval $[0,\pi]$
the boundary conditions (\ref{v4},\ref{v15}) along the space-like boundary
components  can be written as follows:
\begin{eqnarray}
\dot{\varrho}_i\; = \;\dot{\varrho}_{|\partial M_i}\;=\;0\;\;&,&\;\;\;
\dot{\varrho}_f\; =\; \dot{\varrho}_{|\partial M_f}\;=\;0\;\;\;,
\label{td6}\\
x_i\;=\;\widetilde{x}_i \circ \widehat{\gamma}_i[\varrho_i]\;\;\;&,&
\;\;\;x_f\;=\;\widetilde{x}_f \circ
\widehat{\gamma}_f[\varrho_f]\;\;\;,
\nonumber\\
\varphi_i\;=\;\widetilde{\varphi}_i \circ
\widehat{\gamma}_i[\varrho_i]\;\;\;&,&
\;\;\;\varphi_f\;=\;\widetilde{\varphi}_f \circ
\widehat{\gamma}_f[\varrho_f]\;\;\;, \nonumber
\end{eqnarray}
where $\varrho_i = \varrho_{|\partial M_i}, \varrho_f =
\varrho_{|\partial M_f}$,
and the diffeomorphisms
$$
\widehat{\gamma}_i[\varrho_i] : [0,\pi] \longrightarrow
[0,\pi]\;\;\;,
\;\;\;\widehat{\gamma}_f[\varrho_f] :  [0,\pi] \longrightarrow
[0,\pi]\;\;\;
$$
are uniquely determined by the equations
\begin{equation}
\label{td7}
\partial_{\sigma}\widehat{\gamma}_i[\varrho_i]\;\propto
\;\exp {\varrho_i\over2}\;\;\;,
\;\;\;\partial_{\sigma}\widehat{\gamma}_f[\varrho_f]\;\propto
\;\exp {\varrho_f\over2}
\;\;\;.
\end{equation}

Let us note that with the boundary conditions (\ref{td5},\ref{td6})
there is no dynamical degree of freedom  in the metric sector.
Indeed the only classical solutions to the equation (\ref{td3})
satisfying (\ref{td5},\ref{td6}) are $\varrho_{\rm cl} = {\rm const}$.
Inserting these solutions into (\ref{td4a}) and (\ref{td7}) one
can easily check that $\varrho_{\rm cl}$ completely decouples.
The resulting system is determined by the equations of motions
\begin{eqnarray}
 -\ddot{x}^{\mu} + x''^{\mu} &=&0
 \;\;\;,\label{td8}\\
 -\ddot{\varphi} + \varphi'' &=&0
 \;\;\;,\label{td9}\\
 {\alpha\over 2\pi}
\left(  \dot{x}^2 + x'^2 \right)
+ {\beta \over 2\pi}
\left( \dot{\varphi}^2 + \varphi'^2 \right)
-2{\beta\over\pi} \varphi'' &=&0
\;\;\;,            \label{td10}
\end{eqnarray}
and the constraints
\begin{eqnarray}
{\alpha\over\pi}\dot{x}x' + {\beta\over\pi}\dot{\varphi}\varphi'
 - 2{\beta\over\pi}\dot{\varphi}' &=&0
\;\;\;,\label{td11}\\
\int\limits_0^{\pi} d\sigma \dot{\varphi} &=&0\;\;\;.
\label{td12}
\end{eqnarray}

The phase-space analysis of the system above is straightforward.
The non vanishing Poisson bracket relations are
\begin{eqnarray}
\nonumber
\{ p^{\mu}(\sigma),x^{\nu}(\sigma') \} &=& \eta^{\mu\nu} \delta
(\sigma-
\sigma')\;\;\;,\\
\label{td13}
\{ \omega(\sigma), \varphi(\sigma') \} &=&  \delta (\sigma-
\sigma')\;\;\;.
\end{eqnarray}
The  Hamiltonian generating the inner time evolution of
the system
$$
\widetilde{H} \equiv \int\limits_0^{\pi} d\sigma
            {1\over 2}
\left( {\pi\over \alpha} p^2 +  {\alpha\over\pi} x'^2
+
 {\pi\over \beta} \omega^2 + {\beta\over\pi} \varphi'^2  \right)\;\;\;,
$$
leads to  Hamilton's equations
\begin{eqnarray*}
\dot{f} &=& \{\widetilde{H},f\}\;\;\;,\\
\dot{x}^{\mu}(\sigma) &=& {\pi\over \alpha} p^{\mu}(\sigma)\;\;\;,
\;\;\;
\dot{\varphi}(\sigma) \;=\; {\pi\over \beta} \omega(\sigma)\;\;\;,
\\
\dot{p}^{\mu}( \sigma) &=&  {\alpha\over\pi} {x^{\mu}}''(\sigma)\;\;\;,
\;\;\;
\dot{\omega}(\sigma) \;=\;  {\beta\over\pi} \varphi''(\sigma)\;\;\;.
\end{eqnarray*}
The phase-space constraints are given by
\begin{eqnarray*}
H(\sigma)& \equiv& {\pi\over 2\alpha} p^2 + {\alpha \over 2\pi}x'^2 +
            {\pi\over 2\beta}\omega^2 + {\beta \over 2\pi} \varphi'^2
            - 2{\beta\over\pi} \varphi'' \;=\;0
           \;\;\;,\\
V(\sigma) &\equiv& p\cdot x +
\omega\varphi' - 2\omega'   \;=\;0            \;\;\;,\\
\omega_0& \equiv& {1\over\pi}
\int\limits_0^{\pi} d\sigma \omega\;=\;0 \;\;\;.
\end{eqnarray*}
For any
canonical variable $Z = x^{\mu}, p^{\mu}, \varphi, \omega$
and
for $H(\sigma), V(\sigma)$ we define
the mode expansions
$$
Z(\sigma)\; = \;\sum\limits_{n=0}^{\infty} Z_n\cos n\sigma
\;\;\;\;,\;\;\;\;
H(\sigma)\;=\;\sum\limits_{n=0}^{\infty} H_n\cos n\sigma
\;\;\;\;,\;\;\;\;
V(\sigma)\;=\;\sum\limits_{n=1}^{\infty} V_n\sin n\sigma \;\;\;.
$$
In order to simplify the analysis of the algebra of constraints $H_k,
V_k,
\omega_0$  we introduce
\begin{eqnarray*}
L_0 &\equiv &\pi H_0\;=\;\widetilde{H}\;\;\;\;,\;\;\;\;
L_{\pm k} \;\equiv\;
{\pi\over 2} (H_k \pm i V_k)\;\;\;,\;\;\;k>0\;\;\;,\\
\alpha^{\mu}_0 &\equiv& {\pi\over \sqrt{\alpha}}
p^{\mu}_0\;\;\;\;,\;\;\;\;
\alpha^{\mu}_{\pm k} \;\equiv\; {1\over 2}\left(
 {\pi\over \sqrt{\alpha}} p^{\mu}_k \mp ik\sqrt{\alpha}
x^{\mu}_k\right)\;\;\;,
 \;\;\;k>0\;\;\;,\\
 \beta_0 &\equiv& {\pi\over \sqrt{\beta}} \omega_0\;\;\;\;,\;\;\;\;
\beta_{\pm k} \equiv {1\over 2}\left(
 {\pi\over \sqrt{\beta}} \omega_k \mp ik\sqrt{\beta} \varphi_k\right)\;\;\;,
 \;\;\;k>0\;\;\;.
 \end{eqnarray*}
  By straightforward calculations one gets
\begin{equation}
\label{td14}
L_k = L^x_k + L^{\varphi}_k \;\;\;,
\end{equation}
where
$$
L^x_k\; \equiv \; {1\over 2} \sum\limits_{- \infty}^{+\infty}
\alpha_{-n}\cdot \alpha_{k+n}\;\;\;,\;\;\;
L^{\varphi}_k\; \equiv \; {1\over 2} \sum\limits_{- \infty}^{+\infty}
\beta_{-n}\cdot \beta_{k+n}  \; -2\sqrt{\beta}ik\beta_k
\;\;\;.
$$
The Poisson brackets (\ref{td13}) imply the relations
\begin{eqnarray}
\{ \alpha^{\mu}_m,\alpha^{\nu}_n \} &=& im \eta^{\mu \nu}\delta_{m,-n}
\;\;\;,\nonumber\\
\nonumber
\{L^x_m,\alpha^{\nu}_n\} &=& -in\alpha^{\mu}_{m+n}
\;\;\;,\\
\{L^x_m,L^x_n\} &=& i(m-n)L^x_{m+n}\;\;\;,\nonumber\\
\{\beta_m,\beta_n\} &=& im\delta_{m,-n}
\;\;\;,\nonumber\\
\nonumber
\{L^{\varphi}_m,\beta_n\} &=& -in\beta_{m+n} - 2
m^2\sqrt{\beta}\delta_{m,-n}
\;\;\;,\\
\{L^{\varphi}_m,L^{\varphi}_n\} &=& i(m-n)L^{\varphi}_{m+n} - 4\beta i m^3
\delta_{m,-n}\;\;\;,\nonumber
\end{eqnarray}
Calculating the algebra of constraints one gets
\begin{eqnarray}
\label{td16}
\{ L_m,L_n \}& = &i(m-
n)L_{n+m} -4\beta i m^3\;\;\;,\\
\{ L_m,\beta_0 \} &=& \;0\;\;\;. \nonumber
\end{eqnarray}
An important property of the algebra above is that $\left\{
L_m \right\}_{m > 0}$ is a family of second kind
constraints.

We close this section by the
formulae for the conserved  charges related
to the global Poincare symmetry. Using the Noether method one gets
the conserved currents:
\begin{eqnarray*}
j^a_{\mu} & = & {\alpha\over\pi} \sqrt{-g} g^{ab}\partial_b x_{\mu}\;\;\;,\\
j^a_{\mu \nu} &=& {\alpha\over\pi}
 \sqrt{-g} g^{ab} \partial_b x_{[\mu}x_{\nu]}
\;\;\;.
\end{eqnarray*}
In the phase space the total energy momentum of the string
is given by
$$
P^{\mu} = \int\limits_{\Gamma} n^aj^{\mu}_a \, eds=
\int\limits_0^{\pi} p^{\mu}(\sigma)\,d\sigma = \pi p^{\mu}_0\;\;\;,
$$
where $\Gamma$ is an arbitrary curve connecting opposite sides of the
strip of parameters $\sigma,\tau$.
The total angular momentum of the string reads
\begin{eqnarray}
M_{\mu\nu}& =&\int\limits_{\Gamma} n_aj_{\mu\nu}^a =
\int\limits_0^{\pi}( p^{\mu}(\sigma)x^{\nu}(\sigma)-
p^{\nu}(\sigma)x^{\mu}(\sigma) )\,d\sigma  \;\;\;,\nonumber\\
&= &P^{\mu}x_0^{\nu} - P^{\nu}x_0^{\mu} + i\sum\limits_{n=1}^{\infty}
(\alpha^{\mu}_{-n} \alpha^{\nu}_{n} - \alpha^{\nu}_{-n}
\alpha^{\mu}_{n})
\;\;\;.\nonumber
\end{eqnarray}

\section{Target space dynamics}

In this subsection we shall discuss the space-time interpretation
of classical solutions of the massive string model. For that purpose let us
briefly recall the assumptions concerning the target space behavior of
classical solutions in the Nambu-Goto and the BDHP string models.
The Nambu-Goto string in the orthogonal gauge and the BDHP string
in the conformal gauge are determined by the same set of equations
\begin{eqnarray}
\label{cs1}
-\ddot{x}^{\mu} + {x^{\mu}}'' &=& 0\;\;\;,\\
\label{cs2}
\dot{x}^2 + {x'}^2 &=&0\;\;\;,\\
\label{cs3}
\dot{x}\cdot x' &=&0\;\;\;,
\end{eqnarray}
with the boundary conditions
\begin{equation}
\label{cs4}
{x'}^{\mu}(0,\tau) ={x'}^{\mu}(\pi,\tau)=0\;\;\;.
\end{equation}

A simple consequence of (\ref{cs2}) and (\ref{cs4}) is that the ends of
string move with the speed of light. Another one is that the
string world sheet is time-like i.e.
\begin{equation}
\label{cs5}
\det \partial_ax^{\mu} \partial_bx_{\mu} =
-(\dot{x}^2)^2 \leq 0\;\;\;.
\end{equation}
Note that this is the property one has to assume for all trajectories in
the Nambu-Goto string model. Indeed only for time-like trajectories
the Nambu-Goto action is a real-valued functional and the orthogonal
gauge is well defined \cite{vlvo}. In the BDHP model the action is real and
well defined on trajectories $(g,x)$ with $g$  Lorentzian and
non degenerate on the whole strip of parameters (including boundaries).
Only for such trajectories one can prove the validity of the
conformal gauge. In this model the property (\ref{cs5}) is a
consequence of the constraint equations (\ref{cs2},\ref{cs3}).

In the relativistic theory of classical point-like particles
the causality principle is formulated as the condition for the
energy-momentum of the particle to be time-like. In the case
of relativistic one-dimensional extended objects the notion of
causal motion is less obvious and depends on the way such objects may
interact with themselves and other classical systems.
In the commonly accepted formulation of
causality in the classical Nambu-Goto and BDHP  models
the string is regarded as a collection of points which may
individually interact. One says that a string trajectory is causal
if there exists a parameterization such that all points of the
string move with the speed less or equal to the speed of light \cite{he,se}.
Such property of string trajectory we shall call micro-causality.
If we assume that classical strings may
interact only as a whole the causality principle is much less
restrictive - it requires
the spectral condition
$$
P^2 \leq 0\;\;\;,
$$
where $P^{\mu}$ is the total energy-momentum of the string.
Trajectories satisfying the spectral condition above will be
called macro-causal.

The notion of micro-causality plays an important role in the classical
Nambu-Goto and BDHP string models. First of all
for micro-causal trajectories one can
show the validity of the light-cone gauge which in order allows to
find all micro-causal solutions
of the system (\ref{cs1}--\ref{cs4}) \cite{he,se}.
Secondly, the
micro-causality implies the spectral condition for classical
solutions \cite{he} . The inverse implication does not seem to be true
but we do not know  any macro-causal solution of the Nambu-Goto
model which is not
micro-causal. What can be easily shown is that the system
(\ref{cs1}--\ref{cs4}) admits tachyonic motions. A simple example
in the three-dimensional target space is given by
\begin{eqnarray*}
t(\sigma,\tau)&=& \cos(\tau + \sigma) +
\cos(\tau - \sigma)\;\;\;,\\
x(\sigma,\tau)&=&{1\over 2} (\sin^2(\tau + \sigma)
+\sin^2(\tau - \sigma)) \;\;\;,\\
y(\sigma,\tau)&=& \tau - {1\over 4}(\sin 2(\tau +\sigma) +
\sin 2(\tau -\sigma))\;\;\;.
\end{eqnarray*}

To conclude this brief discussion of  causality
in the classical Nambu-Goto model let us
mention that it is related to a rather subtle structure of the phase
space which has no impact on quantum string models obtained by
the covariant quantization techniques.
 As far as the system (\ref{cs1}--\ref{cs4})
is regarded as a two-dimensional $\sigma$-model its reduced
phase space can be identified with the space
 of all classical solutions.
If we however consider the same system as a classical
string model  the corresponding phase space is "smaller"
and consists of
solutions satisfying some causality assumptions. In the so
called old covariant approach   the "big" phase space
is quantized. It turns out however that the quantum
spectral condition
automatically appears due to the properties of
the Fock space holomorphic representation (the tachyonic
ground state
appearing there is of a different origin and has nothing to do
with  classical tachyonic motions).
As we shall see the same
phenomenon takes place in the covariant quantization of
the massive string model.

Let us now turn to the classical solutions of the model
derived in the
previous subsections.
The general solution to the equations
of motion (\ref{td8},\ref{td9}) satisfying
the boundary conditions (\ref{td5}) can be written in the following
standard form
\begin{eqnarray*}
x^{\mu}(\sigma,\tau) &=&
{1\over 2} (f^{\mu}(\tau+\sigma) + f^{\mu}(\tau-\sigma))\;\;\;,\\
\varphi(\sigma,\tau) &=&
{1\over 2} (h(\tau +\sigma) + h(\tau -\sigma) )\;\;\;,
\end{eqnarray*}
where $f^{\mu}(z),h(z)$ are arbitrary functions such that
$$
f^{\mu}(z + 2\pi) = f^{\mu}(z) + {2\pi\over \alpha}P^{\mu}\;\;\;,\;\;\;
h(z +2\pi) = h(z) + {2\pi^2\over \beta}\omega_0\;\;\;,
$$
and $P^{\mu}, \omega_0$ are the total energy-momentum of the string
and the zero
mode of the Liouville momentum, respectively.

Inserting the general solution  to the constraint equations
(\ref{td10}--\ref{td12})
one gets the  equation for the functions $f^{\mu}, h$
\begin{equation}
\label{cs6}
      {\alpha\over 2} f'^2 + {\beta\over 2}h'^2 -2\beta h''=0\;\;\;,
\end{equation}
and the periodicity condition $h(z+ 2\pi) = h(z)$.
The equation above can be seen as the
Hill's equation \cite{se}
\begin{equation}
\label{cs7}
8\beta H'' = -{\alpha\over 2} f'^2 H
\end{equation}
with respect to the function $H= \exp(-{1\over 4}h)$.  The problem is
to
find conditions for the function $ f'^2$
under which the equation above admits strictly positive periodic
solutions.
Since we are not aware of any simple method to solve this problem we
shall
restrict ourselves in the present paper to a simple ansatz giving a
subclass of solutions, large enough to exhibit peculiar features
of the system at hand.

We start with the discussion of micro-causal
solutions to the equation
(\ref{cs6}). For such solutions the ends of string cannot move with
the speed greater than the speed of light and one has
$f'^2(z) \leq 0$.
Then by the constraint equation (\ref{cs7})
${H''} \geq 0$,
and the function $H'$ is monotonic. Since  $H'$ is  periodic  this is
possible only for $h={\rm const}$. One obtains a rather surprising
conclusion that the micro-causality does not allow for the Liouville
excitations in the classical solutions of the
massive string model. Moreover the micro-causal solutions of this model
 precisely coincide with the micro-causal solutions
of the Nambu-Goto and BDHP models. It means that with the micro-causality
principle imposed all three classical string models are identical.

Let us now  consider  macro-causal solutions. A large subclass of such
solution can be obtained by the following light-cone ansatz.
In the Minkowski target space we introduce the light-cone
coordinates
$$
x^{\pm} = {1\over \sqrt{2}} (x^0 \pm x^{d-1})
$$
in a two-dimensional time-like subspace and the transverse
coordinates $\left\{x^i_{\rm tr}\right\}_{i=1}^{d-2}$ in its orthogonal
complement. In these coordinates the Lorentzian scalar product
reads
$$
x\cdot y = - x^+y^- -x^-y^+ + x_{\rm tr}^2\;\;\;.
$$
For $f^+$ of the form
\begin{equation}
\label{cs8}
f^+(z) =  {P^+\over \alpha}z + c^+\;\;\;,
\end{equation}
 the constraint equation (\ref{cs6})
can be solved with respect to $f^-$
\begin{equation}
\label{cs9}
f^-(z) =  {1\over P^+} \int\limits_0^z
\left({\alpha\over 2} {f'}_{\rm tr}^2 + {\beta\over 2}{h'}^2 -2\beta h''
\right)
\,dz'\;+\;c^-\;\;\;.
\end{equation}
It follows that the collection $(f^+,f^-,f_{\rm tr},h)$
where $f_{\rm tr},h$ are arbitrary functions satisfying
the periodicity conditions
$$
f_{\rm tr}^i(z + 2\pi) = f_{\rm tr}^i(z) +
{2\pi\over \alpha}P_{\rm tr}^i\;\;\;,\;\;\;
h(z +2\pi) = h(z) \;\;\;,
$$
and  $f^+,f^-$ are given by the formulae (\ref{cs8},\ref{cs9}),
is a classical solution of the massive string.
Solutions obtained in this way are macro-causal. Indeed,
calculating $2P^+P^-$ one gets
\begin{eqnarray*}
2P^+P^- &=& {\alpha\over  \pi} \int\limits_{-\pi}^{\pi} \left(
{\alpha\over 2} {f'}_{\rm tr}^2 + {\beta\over 2}{h'}^2 -2\beta h''
\right)
\,dz \\
&=& P^2_{\rm tr} +
{\alpha\over \pi} \int\limits_{-\pi}^{\pi} \left(
{\alpha\over 2}
\left(f'_{\rm tr}-{P_{\rm tr}\over\alpha}\right)^2
+ {\beta\over 2}{h'}^2
\right)
\,dz\;\;\;,
\end{eqnarray*}
which implies
the spectral condition
$
P^2=-2P^+P^- + P^2_{\rm tr} \leq 0\;\;\;.
$

Let us note that what we really need to
linearize the constraint equation
(\ref{cs6}) in the light-cone coordinates is the condition for
the function $(f^+)'$ to be strictly positive. With this condition
satisfied one can calculate $f^-$ in terms of $f^+,f_{\rm tr},$
and $h$:
$$
f^-(z) = \int\limits_0^z
{{\alpha\over 2} {f'}_{\rm tr}^2 + {\beta\over 2}{h'}^2 -2\beta h''
\over \alpha {f^+}'}
\,dz'\;+\;c^-\;\;\;.
$$
It is convenient to formulate the ansatz above in a slightly
different way. Let us first observe that any function $f^+$
with strictly positive and periodic derivative can be expressed
in the form
$$
f^+(z) = {P^+\over \alpha} \gamma(z) + c^+\;\;\;,
$$
where $\gamma$ is an orientation preserving diffeomorphism of the real
line such that
\begin{equation}
\label{cs10}
\gamma(2k\pi) = 2k\pi\;\;\;{\rm for} \;\;\;k\in {\bf Z}\;\;\;,
\end{equation}
and
\begin{equation}
\label{cs11}
\gamma'(z+ 2\pi) = \gamma'(z)\;\;\;.
\end{equation}
By explicit calculation one can show that for any diffeomorphism
$\gamma$ satisfying (\ref{cs10}) and (\ref{cs11}), and for any
light-cone solution
$(f^+={P^+\over \alpha}z + c^+, f^-,f_{\rm tr}, h)$
the collection
\begin{equation}
\label{cs12}
(f^+\circ\gamma,f^-\circ \gamma + \widehat{\gamma},
f_{\rm tr}\circ\gamma, h\circ\gamma + 2\log \gamma')\;\;\;,
\end{equation}
where
$$
\widehat{\gamma}(z) =
{2\beta\over P^+} \int\limits_0^z
\left( 3{{\gamma''}^2 \over {\gamma'}^3} - 2{\gamma'''\over {\gamma'}^2}
\right)\,dz'
$$
is a new solution of (\ref{cs6}) with required periodicity properties.

Note that except of the non-homogeneous part $\widehat{\gamma}$
in the expression for the new $f^-$, the formula (\ref{cs12})
is the  transformation rule for classical variables with respect
to the conformal transformations. The appearance of the non-homogeneous
term
$\widehat{\gamma}$ results from the fact that the constraints
of classical massive string are not conformally invariant.

Calculating $2P^+P^-$ for the deformed light-cone solution (\ref{cs12})
one gets
$$
2P^+P^-= P^2_{\rm tr} +
{\alpha\over \pi} \int\limits_{-\pi}^{\pi} \left(
{\alpha\over 2}
\left(f'_{\rm tr}-{P_{\rm tr}\over\alpha}\right)^2
+ {\beta\over 2}{h'}^2
\right)
\,dz\;-\;{2\beta\over P^+} \int\limits_{-\pi}^{\pi}
{{\gamma''}^2 \over {\gamma'}^3}\,dz\;\;\;.
$$
It follows  that the transformation (\ref{cs12})
may lead to tachyonic solutions. Indeed a simple
example of a tachyonic string motion is given by
$f_{\rm tr} = h =0$ and $\gamma \neq {\rm id}$.

In order to illustrate some features of the macro-causal solutions
let us consider a particular case of the light-cone solution in
3-dimensional Minkowski space time given by the functions
\begin{eqnarray*}
f^+(z) &=& {P^+\over \alpha}z\;\;\;,\\
f^x_{\rm tr}(z) &=& a \cos mz\;\;\;,\\
h(z)&=& b \cos nz\;\;\;.
\end{eqnarray*}
The  corresponding string world sheet is
described by
\begin{eqnarray}
t(\sigma,\tau) &=& \left( {P^+\over \sqrt{2}\alpha} +
       {\alpha a^2 m^2 +\beta b^2 n^2 \over 4\sqrt{2} P^+}
        \right) \tau \nonumber\\
& -& { \alpha a^2 m \over 16\sqrt{2} P^+} \left(
      \sin 2m (\tau +\sigma) + \sin 2m (\tau -\sigma) \right)
\nonumber \\
& -& { \beta b^2 n \over 16\sqrt{2} P^+} \left(
      \sin 2n (\tau +\sigma) + \sin 2n (\tau -\sigma) \right)
\nonumber\\
&+& { \beta b n \over \sqrt{2} P^+}
    \left(
      \sin n (\tau +\sigma) + \sin n (\tau -\sigma) \right) \;\;\;,
\nonumber\\
x(\sigma,\tau)&=& {a\over 2}\left(
      \cos m (\tau +\sigma) + \cos m (\tau -\sigma) \right) \;\;\;,
\label{css}\\
y(\sigma,\tau) &=& \left( {P^+\over \sqrt{2}\alpha} -
       {\alpha a^2 m^2 +\beta b^2 n^2 \over 4\sqrt{2} P^+}
        \right) \tau
\nonumber\\
& +& { \alpha a^2 m \over 16\sqrt{2} P^+} \left(
      \sin 2m (\tau +\sigma) + \sin 2m (\tau -\sigma) \right)
\nonumber\\
& +& { \beta b^2 n \over 16\sqrt{2} P^+} \left(
      \sin 2n (\tau +\sigma) + \sin 2n (\tau -\sigma) \right)
\nonumber \\
&-& { \beta b n \over \sqrt{2} P^+}
    \left(
      \sin n (\tau +\sigma) + \sin n (\tau -\sigma) \right) \;\;\;.
\nonumber
\end{eqnarray}
For the parameters $\alpha=1, \beta={11\over 24}, a=0.6,
b=1, P^+=0.8, n=3,
m=1$ the parametric plot of the string history is presented on Fig.1.
We shall analyze more closely this particular string motion.

First of all, as one could expect from the  discussion above, some
parts of the string oscillate with the speed greater than the speed
of light. In consequence for the amplitudes of this oscillations
high enough one has several disjoint pieces of the string
 on  a fixed target
time hyperplane. The configuration of the string at the time $t=T$
corresponding to the upper bound of the plot on Fig.1 is
shown on Fig.2.a. This strange behavior of string trajectory is
in apparent contradiction even with the weaker principle of
macro-causality. What saves the day
is that all classical
macro-observables of the string are concentrated on its one "material"
piece. For all other "ghost" pieces  the total
energy-momentum and the angular momentum are zero. This can be easily
seen from the structure of the equal time lines in the space of
parameters i.e. curves $\gamma(s)=(\sigma(s),\tau(s))$
determined by the implicit
equation
$$
t(\sigma(s),\tau(s)) = T\;\;\;.
$$
For the configuration of Fig.2.a. these lines are drown on Fig.2.b..

It is clear that calculating appropriate line integrals of conserved
currents one gets nonzero result only for the curve $m$ connecting
opposite sides of the strip of parameters. The only "material"
piece of the string is that corresponding to the curve $m$
( the line $M$ on Fig.2.a.). Due to this property the strange
time evolution
of "ghost" pieces of the string (they appear, disappear, split and
join with each other and with the "material" piece) is in a perfect
agreement with the macro-causality principle and the global
conservation laws.

\iffigs
\epsfbox{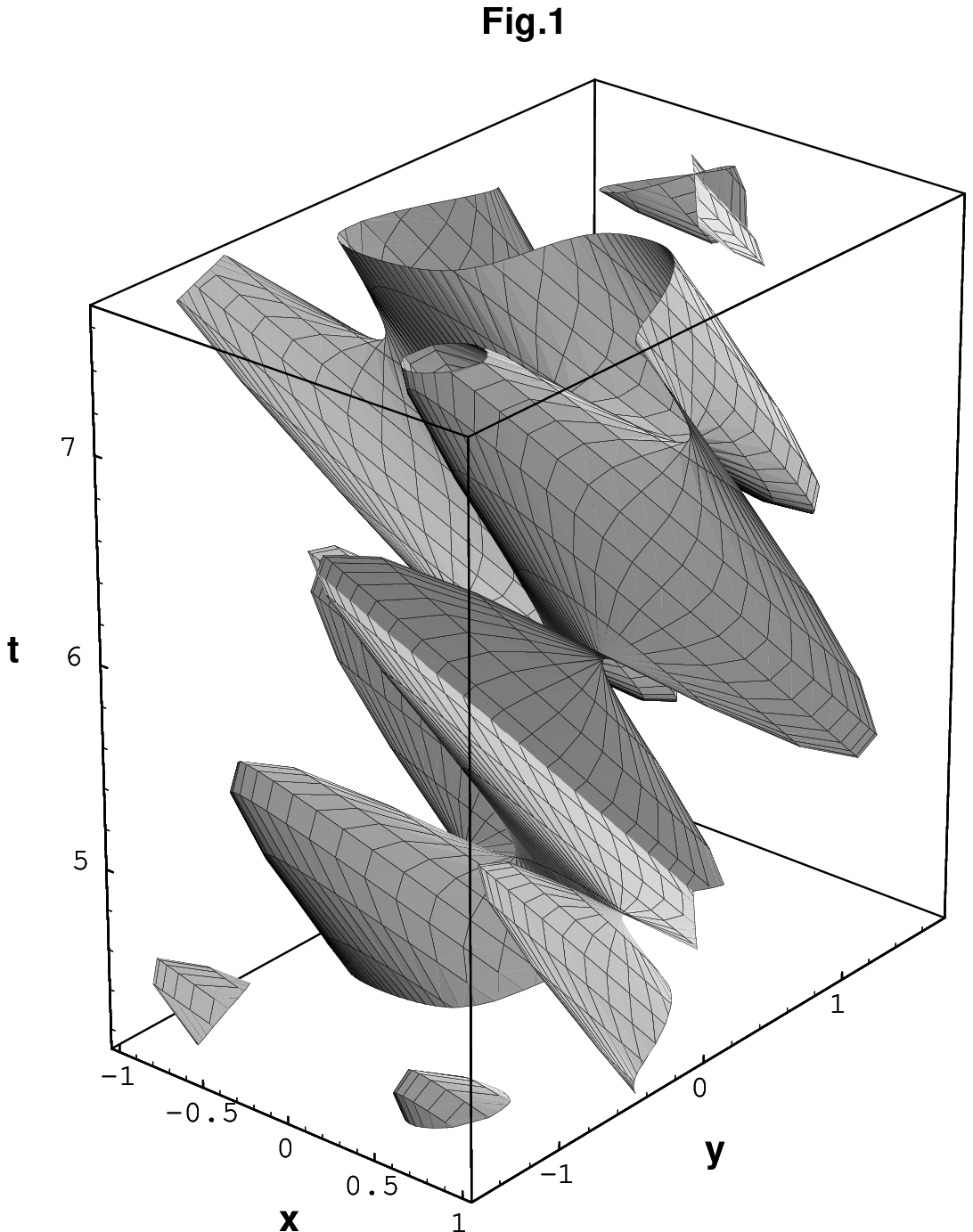}\bigskip\bigskip\bigskip

\epsfbox{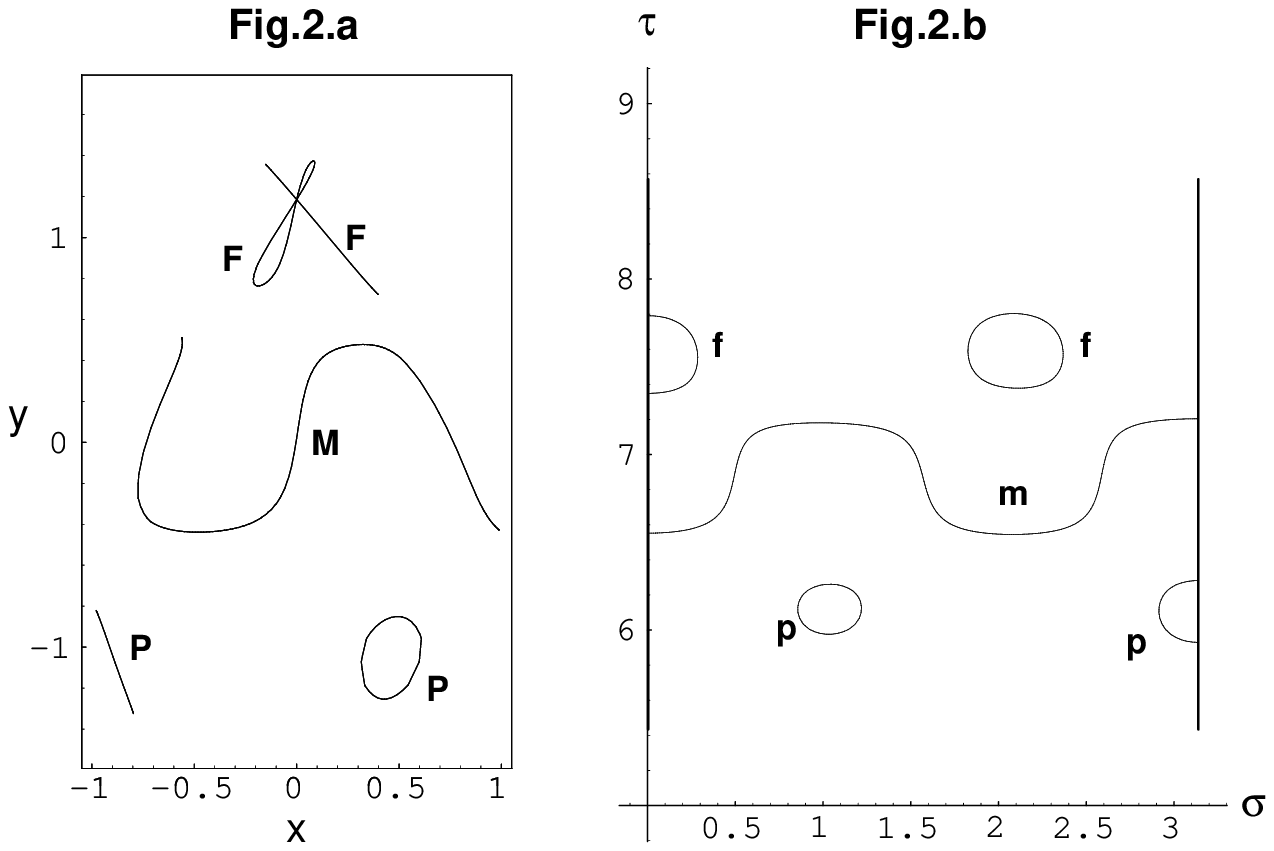} \bigskip
\fi

Another important feature of the solution (\ref{css}),
closely related to the structure
of equal time lines, is that
 the "ghosts" pieces are inessential from
the point of view of the Cauchy problem for the target space
dynamic. Indeed in order to determine the
evolution of the system
it is enough to know "positions" and "velocities"
of all classical variables at each point of the "material" piece
of the string at a fixed target time.
Solving the equations of motions in both time
directions one obtains all "future ghosts" (lines $F$ on Fig.2.a.)
and all "past ghosts" (lines $P$ on Fig.2.a.). It means that
the macro-causality principle allows for a consistent formulation
of target space dynamics, and therefore, for a "stringy" interpretation
of the solution (\ref{css}).

The question arises whether
the discussion of one particular macro-causal solution given above
is valid in general. The main property of the equal time lines required
for the "stringy" interpretation is that for each moment of the target
space time one has only one "material" piece of the string.
Analyzing the geometry of levels of the time component $x^0(\sigma,\tau)$
of a general macro-causal solution one can easily show that this property
holds
except instants corresponding to saddle points of  $x^0(\sigma,\tau)$.
 At these moments  "ghost" pieces join or separate from   the "material"
 one so
the identification  of the "material" piece is ambiguous. This ambiguity
however does not lead to any ambiguity in the Cauchy problem --
all possible choices of  "material" piece at an "interaction time"
lead to the same string trajectory and the same values of macro-observables.

It follows from the considerations above that
one can provide a consistent "stringy"
interpretation for all macro-causal solutions of the massive string
model.
The space of these solutions can be regarded as
the reduced phase space of
the massive string. It is essentially bigger that the reduced phase space of
the Nambu-Goto and the BDHP string models.
It seems that the generalized
light cone ansatz  provides a good local parameterization of
the reduced phase space  but the analysis of singularities of this
parameterizations is a  difficult open problem.
Since solving constraints before quantization
seems to be a very difficult task
the only available method is to represent  the algebra of constraints
on the quantum level.  In the next section we shall discuss
the covariant quantization of the "big" phase space.
As in the case of the Nambu-Goto string
the quantum spectral condition automatically appears.

\section{First quantized massive string}

In these section we  discuss the covariant operator quantization
of the massive string. Following standard   prescriptions of
string theory \cite{re,sch} we start with  the CCR algebra
\begin{eqnarray}
[\alpha^{\mu}_m,\alpha^{\nu}_n] &=& m \eta^{\mu\nu} \delta_{m,-n}\;\;\;,
\nonumber \\
{ [ \beta_m , \beta_n ]}  & = & m \delta_{m,-n} \;\;\;,
\;\;\;m,n\neq 0\;\;\;,
\label{q1}\\
{[P_{\mu},x_0^{\nu}]}& =& - i \delta^{\mu}_{\nu}\;\;\;.
\nonumber
\end{eqnarray}
supplemented by  the conjugation properties
$$
\left(P_{\mu}\right)^+=P_{\mu}\;\;\;,\;\;\;
\left(x_0^{\nu}\right)^+=x_0^{\nu}\;\;\;,\;\;\;
\left( \alpha^{\mu}_m \right)^+ = \alpha^{\mu}_{-m}\;\;\;,\;\;\;
\left( \beta_m \right)^+ = \beta_{-m}\;\;\;.
$$
The space of states is a direct sum of the Fock spaces
$F_p$ along $d$-dimensional spectrum of momentum operators:
$$
{\cal H} = \int d^dp \, F_p\;\;\;.
$$
In each $F_p$ there is a unique vacuum state
$\Omega_p$
satisfying
$$
\alpha^{\mu}_m \Omega_p = \beta_m \Omega_p = 0\;\;\;,\;\;\;m>0\;\;\;;
\;\;\;P_{\mu} \Omega_p = p_{\mu} \Omega_p\;\;\;.
$$
Despite of the conclusions from the classical analysis we do not
impose any condition on the square of momentum.

Because of the presence of the Lorentzian metric in (\ref{q1}),
the scalar product generated on ${\cal H}$ is not positive.
For this reason  we shall consider the Fock part of ${\cal H}$
in a purely algebraic way i.e. we assume that all vectors in ${\cal
H}$
are given by polynomials in creation operators.

 In order to define the constraints  operators (\ref{td14})
we introduce the standard normal ordering of quadratic expressions:
\begin{equation}
\label{q2}
:\alpha^{\mu}_m \alpha^{\nu}_n : = \left\{ \parbox{100pt}{
$\alpha^{\mu}_m \alpha^{\nu}_n\;\;\;m<0$\\
$\alpha^{\nu}_n \alpha^{\mu}_m\;\;\;m\geq 0$} \right.,
\end{equation}
with similar rules for $\beta_m\beta_n$ and mixed products.
With this definition the action of the normally ordered constraints
operators
\begin{equation}
\label{q3}
L_n =  {1\over 2} \sum\limits_{m=- \infty}^{+\infty}
:\alpha_{-m}\cdot \alpha_{n+m}:
+ {1\over 2} \sum\limits_{m=- \infty}^{+\infty}
:\beta_{-m} \beta_{n+m}:  \; -2\sqrt{\beta} ik\beta_k +
2\beta\delta_{n,0}
\;\;\;,
\end{equation}
on the vacuum state is well defined. The action
on states from ${\cal H}$ is then
uniquely determined by the commutation relations
with excitation operators:
\begin{eqnarray}
[L_n,\alpha^{\mu}_m] &=& -m\alpha^{\mu}_{m+n}\;\;\;,
\label{q4}\\
{[ L_n,   \beta_m ]} & = & -m\beta_{m+n} + 2in^2 \sqrt{\beta}
\delta_{n,-m}\;\;\;.
\nonumber
\end{eqnarray}
Due to the normal ordering the central term of the algebra of
constraints
gets shifted by ${1\over 12}(d+1)$
$$
[L_n,L_m] = (n-m)L_{n+m} + {1\over 12}(d+1 +
48\beta)(n^3 -n) \delta_{n,-m}\;\;\;.
$$
The physical subspace ${\cal H}_{\rm phy} \subset {\cal H}$ is defined
as the set of all vectors $\psi$ satisfying
$$
(L_n - \delta_{n,0} a_0) \psi =0\;\;\;.
$$
The parameter $a_0$ in the conditions above is left
arbitrary at the moment. It's value will be restricted by the
no-ghost theorem.

In this representation  of the quantum theory
the algebra of Poincare charges is realized by
the translation operators $P^{\mu}$
and the Lorentz generators:
$$
M^{\mu\nu} = (P^{\mu}x_0^{\nu} -
P^{\nu}x_0^{\mu}) +
i\sum\limits_{n>0}(\alpha^{\mu}_{-n}\alpha^{\nu}_n
- \alpha^{\nu}_{-n}\alpha^{\mu}_n )\;\;\;.
$$

 Following the DDF approach \cite{ddf}
 we introduce the formal operator series
\begin{eqnarray*}
X^{\mu}(\theta) & = & q^{\mu}_0 + \alpha^{\mu}_0 \theta +
\sum\limits_{m\neq 0} {i\over m} \alpha^{\mu}_m {\rm e}^{-im\theta}
\;\;\;,\\
\Phi(\theta) &=&
\sum\limits_{ m\neq 0} {i\over m}\beta_m {\rm e}^{-im\theta}
\;\;\;,\\
P^{\mu}(\theta) &=& (X^{\mu})'(\theta) \;=\;
\sum\limits_{m=-\infty}^{\infty} \alpha^{\mu}_m {\rm e}^{-im\theta}
\;\;\;,\\
\Pi(\theta) &=& \Phi'(\theta) \;=\;
\sum\limits_{ m\neq 0} \beta_m {\rm e}^{-im\theta}\;\;\;,
\end{eqnarray*}
where $q_0^{\mu} = \sqrt{\alpha}x_0^{\mu}$ is
the operator canonically conjugate to $\alpha_0$ and $\theta$ is a real
parameter.

{}From (\ref{q1}) one gets
\begin{eqnarray}
[X^{\mu}(\theta),X^{\nu}(\theta')] &=& -2\pi i \eta^{\mu\nu}
\epsilon(\theta -\theta')\;\;\;,
\nonumber\\
{[\Phi(\theta),\Phi(\theta')]} &=& -2\pi i (\epsilon(\theta -\theta')
+(\theta' - \theta))\;\;\;,
\nonumber\\
{[P_{\mu}(\theta),X^{\nu}(\theta')]} &=& -2\pi i \delta_{\mu}^{\nu}
\delta(\theta -\theta')\;\;\;,
\nonumber\\
{[\Pi(\theta),\Phi(\theta')]} &=& -2\pi i (\delta(\theta -\theta')
-1)\;\;\;,
\nonumber\\
{[P_{\mu}(\theta),P_{\nu}(\theta')]} &=& -2\pi i \eta_{\mu\nu}
\delta'(\theta -\theta')\;\;\;,
\nonumber\\
{[\Pi(\theta),\Pi(\theta')]} &=& -2\pi i \delta'(\theta -\theta')
\;\;\;,\nonumber
\end{eqnarray}
Commutation relations with constraint operators follow from
(\ref{q4})
\begin{eqnarray}
{[L_m,X^{\mu}(\theta)]} & = & -iP^{\mu}(\theta){\rm e}^{im\theta}
\;\;\;,\nonumber\\
{[L_m,\Phi(\theta)]} & = & -i\Pi(\theta){\rm e}^{im\theta}
+2m\sqrt{\beta} {\rm e}^{im\theta}
\;\;\;,
\label{q5}\\
{[L_m,P_{\mu}(\theta)]} & = &
-i{d\over d \theta }(P_{\mu}(\theta){\rm e}^{im\theta})
\;\;\;,\nonumber\\
{[L_m,\Pi(\theta)]} & = & -i{d\over d \theta}(\Pi(\theta){\rm
e}^{im\theta})
+ 2im^2\sqrt{\beta} {\rm e}^{im\theta}
\;\;\;.\nonumber
\end{eqnarray}

For a  fixed light-like vector $k$ $(k^2 =0)$ we define a
basis
of
DDF operators as follows.
Let $k'$ be a
light-like vector satisfying the condition $k\cdot k' = -1$
and  $\left\{ e_i \right\}_{i=1}^{i=d-2}$ -- a basis in the Euclidean
subspace orthogonal to both $k$ and $k'$. We  define
$d-2$ families of the transverse operators \cite{ddf}
\begin{equation}
\label{q6}
A_m^i (k) = {1\over 2\pi} \int\limits_0^{2\pi}
d\theta :e_i \cdot P(\theta)
{\rm e}^{imk\cdot X(\theta)}:\;\;\;,
\end{equation}
and one family of the longitudinal (Brower) vortices \cite{bro}
\begin{equation}
\label{q7}
\widetilde{B}_m (k) = {1\over 2\pi} \int\limits_0^{2\pi}
d\theta :(k'\cdot P(\theta) - {im\over 2}  \log'(k\cdot
P(\theta)))
{\rm e}^{imk\cdot X(\theta)}:\;\;\;.
\end{equation}
In addition we introduce the operator corresponding to the Liouville
degree
of freedom \cite{jame}
\begin{equation}
\label{q8}
C_{m}(k) = {1\over 2\pi} \int\limits_0^{2\pi}
d\theta :(\Phi(\theta) - 2\sqrt{\beta} \log'(k\cdot P(\theta)))
{\rm e}^{imk\cdot X(\theta)}:\;\;\;.
\end{equation}
The primes over logarithmic terms denote derivatives with respect to
$\theta$.

In order to have well defined operators the definition of the
normal ordering  (\ref{q2}) has to be supplemented by the following
rules
\begin{eqnarray*}
:{\rm e}^{imk\cdot X(\theta)}: &=&
{\rm e}^{imk\cdot X_-(\theta)}{\rm e}^{imk\cdot X_0(\theta)}
{\rm e}^{imk\cdot X_+(\theta)}\;\;\;,\\
:\xi\cdot P(\theta)
{\rm e}^{imk\cdot X(\theta)}: &=&
\xi\cdot P_-(\theta){\rm e}^{imk\cdot X(\theta)} +
{\rm e}^{imk\cdot X(\theta)}\xi\cdot P_+(\theta)\;\;\;,
\end{eqnarray*}
where
\begin{eqnarray*}
X_{\pm}(\theta) &=& \pm \sum\limits_{m>0} {i\over m} \alpha_{\pm m}
{\rm e}^{\mp im\theta}\;\;\;,\;\;\;
X_0(\theta) \; =\; q_0 + \alpha_0 \theta\;\;\;,\\
P_{\pm}(\theta) &=& {1\over 2}\alpha_0 + \sum\limits_{m>0}
\alpha_{\pm m} {\rm e}^{\mp im\theta}\;\;\;,\;\;\;
\Pi_{\pm}(\theta) \;=\;  \sum\limits_{m>0}
\beta_{\pm m} {\rm e}^{\mp im\theta}\;\;\;.
\end{eqnarray*}
The logarithmic terms should be understood  as power series expansions
around the eigenvalue of the zero-mode $k\cdot \alpha_0$.
With these prescriptions, the power series present in (\ref{q6}--
\ref{q8})  are reduced to polynomials in excitations
on the subspaces $F_p \subset {\cal H}$
with  $p$ satisfying $k\cdot p = \sqrt{\alpha}$.

Calculating
the algebra of DDF operators (\ref{q6},\ref{q7},\ref{q8})
one gets
\begin{eqnarray*}
[A_m^i(k),A_n^j(k)] & = & m \delta^{ij} \delta_{m,-n}\;\;\;,\\
{[C_m(k),C_n(k)]} &=& m\delta_{m,-n}\;\;\;,\\
{[\widetilde{B}_m(k),\widetilde{B}_n(k)]} & = &
(n-m) \widetilde{B}_{n+m}(k) +
2n^3\delta_{m,-n}
\;\;\;,\\
{[\widetilde{B}_n(k), A_m^i(k) ]} &=& -m A^i_{m+n}(k) \;\;\;,\\
{[\widetilde{B}_n(k), C_m(k)]} & =&
-mC_{n+m}(k)+2in^2\sqrt{\beta} \delta_{n,-m}\;\;\;.
\end{eqnarray*}
This algebra  can be diagonalized by  the following
shift of the Brower vertex \cite{bro}
\begin{equation}
\label{q9}
B_n(k) = \widetilde{B}_n(k) - {\cal L}_n(k) + \delta_{n,0}\;\;\;,
\end{equation}
where
$$
{\cal L}_n(k) = {1\over 2} \sum\limits_{m=-\infty}^{+\infty}
\sum\limits_{i=1}^{d-2} A^i_{-m}(k)A^i_{n+m}(k) +
{1\over 2} \sum\limits_{m=-\infty}^{+\infty}
C_{-m}(k)C_{n+m}(k) + 2in\sqrt{\beta} C_n(k) + 2\beta \delta_{n,0}\;\;\;.
$$
In the new basis $A^i_m(k), B_m(k), C_m(k)$ the only nonzero
commutators are
\begin{eqnarray}
{[A_m^i(k),A_n^j(k)]} & = & m \delta^{ij} \delta_{m,-n}\;\;\;,\nonumber\\
{[B_n(k),B_m(k)]}& = &(n-m)B_{n+m}(k) + {1\over 12}(n^3 -n)
(25-d-48\beta)\delta_{n,-m}
\;\;\;,\label{q10}\\
{[C_m(k),C_n(k)]} &=& m\delta_{m,-n}\;\;\;.\nonumber
\end{eqnarray}

Using  (\ref{q5}) one can find out the commutation
relations
of the DDF operators with the constraints (\ref{q6})
$$
[ L_n, A_m^i(k) ] = [ L_n, B_m(k)] = [ L_n, C_m(k)] = 0\;\;\;,
$$
for all $m,n \in {\bf Z}$.
It follows that acting on  vacuum states $\Omega_p$ such that $k\cdot p
=\sqrt{\alpha}$
 the DDF operators $A_m^i(k), B_m(k), C_m(k)$
generate off-shell physical states
i.e. states satisfying the physical off-shell condition
$$
L_n \Psi =0\;\;\;,\;\;\;n>0\;\;\;.
$$
All states obtained in this way (for  different $k$)
we shall call the DDF states.

As a preparation to  the no-ghost theorem we introduce
(for a fixed light-like vector $k$) the  family of operators
\cite{brgo}
$$
F_m(k) = {1\over 2\pi} \int\limits_0^{2\pi} d\theta\,
:{\rm e}^{imk\cdot X(\theta)}: \;\;\;.
$$
The commutation relations read
\begin{eqnarray}
[A_m^i(k),F_n(k)] &=& [C_m(k),F_n(k)] \;=
\;[F_m(k),F_n(k)]\;=\;0\;\;\;,\nonumber\\
{[B_m(k),F_n(k)]} &=& -nF_{n+m}(k)\;\;\;,
\nonumber
\end{eqnarray}
In contrast to the DDF operators $F_m(k)$ do not commute with the
constraints
\begin{eqnarray}
\label{q11}
[L_m,F_n(k)]& =& -mF_n^m(k)
\;\;\;,\\
F_m^n(k) &=& {1\over 2\pi}
\int\limits_0^{2\pi} d\theta\,{\rm e}^{in\theta}
:{\rm e}^{imk\cdot X(\theta)}: \;\;\;.\nonumber
\end{eqnarray}
For a vacuum state $\Omega_p$ with $p\cdot k=\sqrt{\alpha}$ one has
\begin{eqnarray}
F_{-m}^n(k) \Omega_p &=&0\;\;\;,\;\;\;m>0,n>m\;\;\;,\nonumber\\
F_{-m}^m (k)\Omega_p &=& \Omega_p\;\;\;.
\label{q12}
\end{eqnarray}

Let us  consider the states of the form
\begin{eqnarray}
\Psi_{ ( \{ \overline{a} \},\{b\},\{ c \},\{f\} ) }^{(N)}
&=&
\prod\limits_{i=1}^{d-2}
\left(A^i_{-m_i}\right)^{a^i_{m_i}}\cdot ...
\cdot \left(A^i_{-1}\right)^{a_1^i} \label{q13}\\
& &\cdot\widetilde{B}^{b_n}_{-n}\cdot ...\cdot
\widetilde{B}^{b_1}_{-1} \cdot
C^{c_l}_{-l}\cdot ...\cdot C^{c_1}_{-1} \cdot
F^{f_k}_{-k}\cdot ...\cdot F^{f_1}_{-1}
\Omega_{p+ Nk}\;\;\;,\nonumber
\end{eqnarray}
where $ \{ a^i \},\{b\},\{c\},\{ f \}$ are arbitrary finite sequences
of nonnegative integers such that the eigenvalue
$$
N = \sum\limits_{i=1}^{d-2}
\sum\limits_{r_i=1}^{m_i} r_ia^i_{r_i} + \sum\limits_{s=1}^{n}
 s b_s  +
\sum\limits_{t=1}^{l} t f_t  +\sum\limits_{u=1}^{k} u f_u\;\;\;
$$
of the level operator $R$ is fixed.
The virtue of the operators $F_m(k)$ is that for $N=0,1,2,...$
the states (\ref{q13}) constitute a basis in the
space ${\cal H}_p$. In order to check this property one can look
at the structure of the operators $A^i_{-m}(k),
\widetilde{B}_{-m}(k),  C_{-m}(k), F_{-m}(k)$
in terms of fundamental creation operators:
\begin{eqnarray}
A^i_{-m}(k) &=& \alpha^i_{-m} - \alpha^i_0 k\cdot \alpha_{-m} + {\rm "more"}
\;\;\;,\nonumber\\
\widetilde{B}_{-m}(k) &=& k'\cdot \alpha_{-m} -
( k'.\alpha_0 - {1\over 2}m(m+1))k\cdot \alpha_{-m} + {\rm "more"}
\;\;\;,\label{q14}\\
C_{-m}(k) &=&
\beta_{-m} - 2i\sqrt{\beta} m k\cdot\alpha_{-m} + {\rm "more"}
\;\;\;,\nonumber\\
F_{-m}(k) &=& k\cdot\alpha_{-m} + {\rm "more"}
\;\;\;.\nonumber
\end{eqnarray}
The terms denoted by "more" are of higher order in the creation operators
$\alpha_{-n}, \beta_{-n}$ with $n<m$. The structure of the "leading
terms" in (\ref{q14}) implies that the states (\ref{q13}) are linearly
independent. Counting their number level by level one can show that they
form a basis of ${\cal H}_p$. This statement holds if we replace the
Brower vortices $\widetilde{B}_n(k)$ in the formula (\ref{q13})
 by the shifted ones $B_n(k)$.

Now we proceed to  the  no-ghost theorem for
the massive string. As in  Brower's proof of the no-ghost theorem for
the Nambu-Goto string  \cite{bro}
the problem can be considered in two steps. We first show that the on-shell
DDF states exhaust all physical states. Then using the algebra
of the DDF operators we shall analyze the metric on
the subspace ${\cal H}_{\rm DDF}$.

 We start with the following lemma \bigskip

\noindent{\bf Lemma}
{\it  Let $p\neq 0$. A state $\Psi \in {\cal H}_p$ is an off-shell physical
 state if and only if it is a DDF state.}\bigskip

Since $p\neq 0$ there exists a light-like vector $k$ such that
$k\cdot p=\sqrt{\alpha}$ and one can use
 the states (\ref{q13}) as a basis in ${\cal H}_p$.
It follows that any state $\Psi \in {\cal H}$
can be written as a sum $\Psi = \Psi_{\rm DDF} + \Psi_{\rm F}$, where
$\Psi_{\rm DDF} \in {\cal H}_{\rm DDF}$ and $\Psi_{\rm F}$ contains
nonzero excitations $F_{-m}$. We shall show that the conditions
$$
L_m \Psi = L_m \Psi_{\rm F} =0\;\;\;\; \;n>0\;\;\;,
$$
imply $\Psi_{\rm F} =0$. In the expression for $\Psi_{\rm F}$ in terms of
the basis (\ref{q13}) there is a term containing the operator $F_{-m}(k)$
with maximal $m$ and raised to the maximal power $f_m$:
\begin{eqnarray}
\label{q15}
\Psi_{\rm F} &=& A F^{f_m}_{-m}(k)\cdot ...
\cdot F^{f_1}_{-1}(k)
\Omega_{p + N_Ak}\;\;\;\\
&+&B F^{f'_m}_{-m}(k)\cdot ...\cdot F^{f'_1}_{-1}(k)
\Omega_{p+ N_Bk}\;+\;...\;\;\;,\nonumber
\end{eqnarray}
where $f_m>f'_m$ and $A, B$ contains only DDF creation operators.
Due to (\ref{q11}) and (\ref{q12}) one has
$$
\left(L_1\right)^{f_1}\cdot ... \cdot \left(L_m\right)^{f_m}
\Psi_{\rm F} = \prod\limits_{j=1}^m f_j \,  ! (-j)^{f_j}
A \Omega_{p+ N_Ak}\;\;\;,
$$
and consequently $A\Omega_{p+N_Ak}=0$.
Hence the first term in the expansion
(\ref{q15}) must vanish. Repeating this procedure for next terms we prove
 $\Psi_{\rm F} =0$.

It follows from the Lemma that the space of physical states ${\cal H}_{\rm
ph}$
coincides
with the space of on-shell DDF states or, which is the same, with the
space of DDF states created from physical vacua.

Let $\Omega_p$ with $p\neq 0$
be a physical vacuum i.e.
$p^2 = -2\alpha( 2\beta - a_0)$.
Consider the subspace ${\cal H}_{\rm DDF}(p,k)\subset
{\cal H}_{\rm ph} $ of states generated from $\Omega_p$ by the
DDF operators with some
$k$, $k\cdot p = \sqrt{\alpha}$.
The metric on  ${\cal H}_{\rm DDF}(p,k)$  is
completely determined by the algebra (\ref{q10})
and the hermicity properties of the operators $A_n(k), B_n(k), C_n(k)$.
Due to the diagonal form of the algebra (\ref{q10}) the
space ${\cal H}_{\rm DDF}(p,k)$ is isomorphic to the symmetric tensor product
of the spaces ${\cal H}_{\rm AC}(p,k)$ and ${\cal H}_{\rm B}(p,k)$
generated from $\Omega_p$ by the algebra of $A_n(k),C_n(k)$ and
the algebra of $B_n(k)$ operators, respectively.
Since the metric on ${\cal H}_{\rm AC}(p,k)$
is positive, the metric structure on ${\cal H}_{\rm DDF}(p,k)$ depends
on the metric structure on ${\cal H}_{\rm B}(p,k)$.

  Calculating the action of the operator $B_0(k)$ (\ref{q9})  on
the physical vacuum one gets
$$
B_0 \Omega_p = (1-a_0)\Omega_p\;\;\;.
$$
It follows  that ${\cal H}_{\rm B}(p,k)$ is the Verma module
 ${\cal V}_{c,h}$  of the
Virasoro algebra with the central charge
$$
c = 25 - d - 48\beta\;\;\;,
$$
and the weight
$$
h = 1 - a_0\;\;\;.
$$
In general
the metric on ${\cal V}_{c,h}$ may be degenerate. Taking the
quotient  by the
subspace of null vectors one gets the irreducible highest weight
representation ${\cal H}_{c,h}$ \cite{fefu}.
It is known \cite{frqish,gokeol} that the representation
 ${\cal H}_{c,h}$ is unitary i.e.
the metric on
${\cal H}_{c,h}$ is positively defined, if and only if one of the two
following
conditions is satisfied
$$
c\geq 1 \;\;\;, \;\;\; h\geq 0\;\;\;,
$$
or
$$
c=c_m \;\;\;,\;\;\; h=h_{rs}(m)
 \;\;\;{\rm for}\;\;\;m= 2,3,... ;\;\; 1\leq r\leq m-1;\;\; 1\leq s\leq r\;\;,
$$
where
$$
c_m\equiv 1 - {6\over m(m+1)}
 \;\;\;,\;\;\;h_{rs}(m)\equiv {((m+1)r -m s)^2 - 1 \over 4m(m+1)}\;\;.
$$
 \bigskip
One gets the following

\noindent{\bf Theorem} {\it
The space of physical states in the massive string model is ghost free
if and only if one of the following two conditions is satisfied:
\begin{equation}
\label{q16}
a_0 \leq 1\;\;\;,\;\;\; 0< \beta \leq {24-d\over 48}
\;\;\;;
\end{equation}
or
\begin{equation}
\label{q17}
\beta \;=\;\beta _m\;\;\;,\;\;\;
a_0\;=\;a_{rs}(m)\;\;\;
{\it for}
\;\; m= 2,3,...;\;\;1\leq r\leq m-1;\;\; 1\leq s\leq r\;\;;
\end{equation}
where
$$
\beta _m\; \equiv \;{24-d\over 48} + {1\over 8m(m+1)}
 \;\;\;,\;\;\;
 a_{rs}(m)\;\equiv\;1 - {((m+1)r -m s)^2 - 1 \over 4m(m+1)}\;\;\;.
$$ }

Due to the structure of the operator $L_0$ (\ref{q3}) the physical
mass spectrum of the massive string is bounded from below.
The theorem above implies that there are no excited
tachyonic states in the physical spectrum.
Indeed, for all admissible values of $\beta, a_0$
only the vacuum states with $m^2 = 2\alpha(2\beta -a_0)$
may be tachyonic.

\section{Conclusions}

The main result of the present paper is that in the range of
dimensions $1<d<25$ the action functional
(\ref{v1}) leads to a new  consistent classical and quantum theory
of one-dimensional relativistic extended objects.

Our derivation of the classical model is based on
 a new reparameterization invariant formulation of the
 variational principle. The virtue of this approach is
a clear geometrical interpretation of the classical system and
a simple phase space formulation of the 2-dim dynamics.
 The most interesting
result of this part of the work is the derivation of the constraint
$\omega_0 =0$. In the standard formulation  \cite{marnel,phas}
the origin of this constraint is not clear and one has to introduce
it by hand in
order to remove the unphysical internal degree of freedom.
In the present approach it appears  as a necessary
consistency condition for the variational principle
to be well posed and diffeomorphism invariant.

Let us note that the Euclidean counterpart of
the variational problem formulated in
Section 2 is interesting by its own. One can easily check that
minimal surfaces form a special
subclass of solutions to such problem with $\varphi = 0$. It would be
interesting to find some local and global geometric characterizations of
solutions with $\varphi \neq 0$.

The analysis of the classical causality
given in Section 4 leads to
the following  conclusions.         First of all  if one assumes
the micro-causality principle the classical massive string
model coincides with the Nambu-Goto and the BDHP models.
Secondly
if one admits the spectral condition for the total energy-momentum of string
as a weaker notion of causality then the space of macro-causal solutions
is essentially bigger than the space of micro-causal ons.
 The important point is
 that macro-causal solutions
 still can be given a consistent "stringy" interpretation.
Both results mean that the classical massive string  is in
a way a minimal generalization of the Nambu-Goto and the BDHP
models.

Our discussion of the target space dynamics leaves  some open questions.
First of all it is desirable to have a detailed description of the
reduced phase space of the massive string defined
as a space of all solutions of the constraint equation
(\ref{cs6}) satisfying the spectral condition.
The light cone ansatz which parameterizes almost all micro-causal solutions
is not a good parametrization for  macro-causal ones.
It seems that the generalized light-cone ansatz may
provide at least a local parametrization of the massive string phase space.
A justification of this conjecture and a clarification of the global
geometric structure are still open problems.
Similar questions are also interesting in the case of old string models,
where the difference between the micro- and macro-causal solutions
is unknown.

As was shown in Section 5 the quantum massive string model can be
obtained by covariant quantization techniques.
The main results of this section are the explicit construction of
physical states by suitably modified DDF method and the no-ghost
theorem yielding necessary and sufficient
conditions for a consistent quantum theory.

One of the problems not analyzed in this paper is the structure of
null states.
Although all technical ingredients
required are known from the conformal field theory with the
central charge $0<c\leq 1$ a comprehensive analysis of this point is rather
involved and we restrict ourselves only to few remarks.
First of all for
$0<\beta< {24-d\over 48}$ there are no null states.
In  this range the symmetry structure and the number of dynamical
degrees of freedom of the quantum theory is the same as in the classical
one.

The largest subspace of null states appears for the critical
values  $\beta_c = \beta_2 ={25-d\over 48}$, $a_c =a_{11}(2)=1$.
In this case the model is equivalent to the old FCT string
\cite{ct} and to the
noncritical Polyakov string \cite{jame}
with extra constraint $\omega_0 =0$.
All states generated by the shifted longitudinal operators
$B_n$ are null and decouple from the physical Hilbert space.
The resulting quantum system has effectively one "functional"
degree of freedom less than the classical one. This phenomenon
can be seen as an extra gauge symmetry (anti-anomaly) of the
quantum model. Due to this special structure of null states
one can use the "quantum" light cone gauge to describe the
physical states of the Polyakov noncritical string \cite{jame}.
Let us stress however that since this symmetry is not present in
the classical model the light cone gauge cannot be used to
solve classical constraints.

Another open problem of the quantum massive string is its
spin spectrum. It is given in terms of the
decomposition of the physical Hilbert space into
irreducible unitary representations of the Poincare algebra.
Such decomposition depends on the structure of null states and is
particularly interesting for the discrete series (\ref{q17}).
As a simple  illustration of the problem
let us consider the structure of $SO(d-1)$
multiplet at  first excited level
in the case of the critical   massive string
(i.e. corresponding to $\beta_c = {25-d\over 48}, a_{c}=1 $).

For a given  on-shell vacuum state $\Omega_p$
there exists a Lorentz frame such that  $p = \lambda k -k'
(k\cdot p=1)$ with $\lambda = 1-2\beta$ (for the sake of simplicity
we put $\alpha = 1$). Acting with DDF operators one gets the
set of states
\begin{eqnarray*}
|a^i\rangle &\equiv & A^i_1\Omega_p = \alpha^i_{-1} \Omega_{p-k}\;\;\;,\\
|b\rangle &\equiv & B_1\Omega_p =
( k'\cdot\alpha_{-1} + 2i\sqrt{\beta} \beta_{-1}
+2\beta k\cdot\alpha_{-1} )\Omega_{p-k}\;\;\;,\\
|c\rangle &\equiv & C_1\Omega_p =
( \beta_{-1} -
2i\sqrt{\beta} k\cdot\alpha_{-1} )\Omega_{p-k}\;\;\;.
\end{eqnarray*}
The little group of the massive vector $p-k= -2\beta k - k'$
is generated by
$$
M^{ie} \equiv - i \sum\limits_{n=1}^{\infty} {1\over n} \left(
(e\cdot \alpha_n) \alpha^i_{-n} - (e\cdot \alpha_{-n}) \alpha^i_{n}
\right)\;\;\;,
$$
where $ e = -2\beta k + k'$ is orthogonal to $p-k$.
By simple calculations one checks that
\begin{eqnarray*}
M^{ie}|a^j\rangle &=& \delta^{ij}2\sqrt{\beta} |c\rangle
- i \delta^{ij} |b\rangle\;\;\;,\\
M^{ie}|b\rangle &=& 0\;\;\;,\\
M^{ie}|c\rangle &=& 2\sqrt{\beta} |a^i\rangle \;\;\;.
\end{eqnarray*}
Consequently up to the null longitudinal state $|b\rangle$ the
states $|a^i\rangle , |c\rangle$ form a linear multiplet with
respect to the little group $SO(d-1)$ of the massive vector $p-k$.

In many respects the critical massive string
is especially interesting. First of all it provides
solution to the problem which was our original motivation
-- the  application of standard quantization techniques to
this classical system yields
the noncritical Polyakov string. This relates the critical massive
string with the Polyakov sum over random surfaces in the
range $1<d<25$.
Secondly the critical massive string
has the largest subspace of null states
and the structure of the quantum theory is the same as in
the critical Nambu-Goto and the BDHP string models.
Finally the rescaling gauge symmetry
of the model, for which the assumption of vanishing
cosmological constant is crucial, implies that
the two-dimensional gravity completely decouples.
Note that the last feature holds for all admissible (not necessary
critical) values of $\beta$ and $a_0$ and yields a chance to overcome
the $c=1$ barrier.
All these properties makes the critical massive string a promising
candidate for a consistent interacting string theory
in physical dimensions.
Of course the most interesting  open question   is whether
such theory exists.\bigskip\bigskip

\noindent{\bf Acknowledgements}
We would like to thank Krzysztof Meissner for discussions at  early
stages of this work and
 Przemys{\l}aw Siemion for his patience in
helping  us with  "Mathematica".
One of us (Z.J.) would like  to thank the mathematics department and
research institute (I.R.M.A) in Strasbourg, where  part of this work was
completed, for their  hospitality.

\end{document}